\documentclass{lmcs} 


\usepackage{hyperref}

\usepackage{bm}
\usepackage{mathtools}
\usepackage{amsmath}
\usepackage{amssymb}
\usepackage{amsfonts}
\usepackage{stmaryrd}
\usepackage{booktabs}
\usepackage[sort&compress, numbers, square, comma]{natbib}
\usepackage{caption}
\usepackage{subcaption}
\DeclareCaptionSubType*[arabic]{figure}
\usepackage{enumitem}
\setlist[itemize]{
    leftmargin = 15pt,
    itemsep = 0.5ex,
    parsep = 0.5ex
}
\setlist[enumerate]{
    itemsep = 0.5ex,
    parsep = 0.5ex
}
\setlist[description]{
    leftmargin = !,
    itemsep = 1ex,
    parsep = 1ex,
    font = {\normalfont\sffamily\slshape}
}
\usepackage[ruled, linesnumbered, lined, noend]{algorithm2e}
\SetArgSty{textnormal}
\SetFuncSty{textsc}
\usepackage{bookmark}
\usepackage{tikz}
\usetikzlibrary{
    automata,
    positioning,
    fit,
    arrows,
    arrows.meta,
    calc
}
\tikzset{
    every picture/.style = {
        thick,
        > = stealth,
        initial text =,
        node distance = 2cm
    },
    every state/.style = {
        inner sep = 3pt,
        outer sep = 0pt,
        minimum size = 15pt,
        font = \footnotesize
    },
    every node/.style = {
        font = \footnotesize
    }
}
\usepackage{tikz-cd}
\tikzcdset{
    arrow style = tikz,
    diagrams = {
        > = Stealth,
    }
}

\newcommand{\tn}[1]{\textup{#1}}
\newcommand{\bb}[1]{\mathbb{#1}}
\newcommand{\mc}[1]{\mathcal{#1}}
\newcommand{\ms}[1]{\mathsf{#1}}
\newcommand{\mb}[1]{\mathbf{#1}}

\newcommand{\set}[1]{\left\{ #1 \right\}}
\newcommand{\paren}[1]{\left( #1 \right)}
\newcommand{\sem}[1]{\ensuremath{\left\llbracket #1 \right\rrbracket}}
\newcommand{\brak}[1]{\bm{\left[} #1 \bm{\right]}}
\newcommand{\tup}[1]{\left\langle #1 \right\rangle}
\newcommand{\card}[1]{\left| #1 \right|}

\newcommand{\eqx}{\overset{\scriptscriptstyle{X}}{\sim}}

\newcommand{\decomp}{\tn{\textsc{Decompose}}}

\DeclareMathOperator{\dom}{\mathsf{dom}}
\DeclareMathOperator{\im}{\mathsf{im}}
\DeclarePairedDelimiter\floor{\lfloor}{\rfloor}
\DeclareMathOperator*{\comp}{\bigcirc}

\newcommand{\teal}[1]{{\color{teal}#1}}

\def\cf{{\em cf.}}
\def\ie{{\em i.e.}}

\begin{document}

\title{Composing Copyless Streaming String Transducers}


\author[R.~Alur]{Rajeev Alur\lmcsorcid{0000-0003-1733-7083}}[a]
\author[T.~Dohmen]{Taylor Dohmen\lmcsorcid{0000-0001-5722-4847}}[b]
\author[A.~Trivedi]{Ashutosh Trivedi\lmcsorcid{0000-0001-9346-0126}}[b]

\address{University of Pennsylvania, 3330 Walnut St, Philadelphia, PA 19104 USA}	
\email{alur@seas.upenn.edu}  

\address{University of Colorado, 1111 Engineering Dr, Boulder, CO 80309 USA}	
\email{taylor.dohmen@colorado.edu, ashutosh.trivedi@colorado.edu}  





\begin{abstract}
Streaming string transducers (SSTs) implement string-to-string transformations by reading each input word in a single left-to-right pass while maintaining fragments of potential outputs in a finite set of string variables. 
These variables get updated on transitions of the transducer, where they can be assigned new values described by concatenations of variables and output symbols. 
An SST is called copyless if every update is such that no variable occurs more than once amongst all of the assigned expressions.  
The transformations realized by copyless SSTs coincide with Courcelle's monadic second-order logic graph transducers (MSOTs) when restricted to string graphs. 
Copyless SSTs with nondeterminism are known to be equivalent to nondeterministic MSOTs as well.

MSOTs, both deterministic and nondeterministic, are closed under composition.
Given the equivalence of MSOTs and copyless SSTs, it is easy to see that copyless SSTs are also closed under composition.
The original proof of this fact, however, was based on a direct construction to produce a composite copyless SST from two given copyless SSTs.
A counterexample discovered by Joost Englefriet showed that this construction may produce copyful transducers.
We revisit the original composition constructions for both deterministic and nondeterministic SSTs and show that, although they can introduce copyful updates, the resulting copyful behavior they exhibit is superficial.
To characterize this mild copyful behavior, we define a subclass of copyful SSTs, called diamond-free SSTs, in which two copies of a common variable are never combined in any subsequent assignment.
In order to recover a modified version of the original construction, we provide a method for producing an equivalent copyless SST from any diamond-free copyful SST.
\end{abstract}

\maketitle

\section{Introduction}
\label{sec:intro}
Streaming string transducers (SSTs) are computational models for transformations between formal languages.
Upon reading a symbol, an SST changes state and updates a finite set of string variables, which are initially set to the empty string $\varepsilon$, using concatenation expressions, such as $x := a y b z c$, made up of variables $y,z$ and symbols $a,b,c$ from a designated output alphabet.
The final output of the transducer is determined by a similar combination of variables and symbols, depending on the final state reached after reading an input word.
Deterministic SSTs (DSSTs) \cite{AlurCerny2010,AlurCerny2011} model functions over strings, while nondeterministic SSTs (NSSTs) \cite{AlurDeshmukh2011} are interpreted as relations.

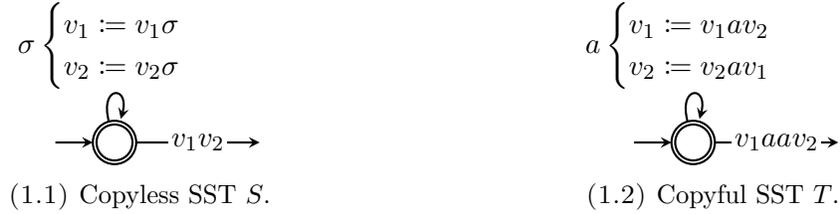
\begin{figure}
	\begin{subfigure}{0.5\linewidth}
		\centering
		\begin{tikzpicture}[every node/.style = {inner sep = 1pt, outer sep = 1pt, minimum size = 0pt, fill = white}]
			\node[state, accepting, initial] (1) {};
			\node (out) [right of = 1] {};

			\path[->] (1) edge[loop above] node {$\sigma \begin{cases} v_1 \coloneq v_1 \sigma \\ v_2 \coloneq v_2\sigma \end{cases}$} (1);
			\path[->] (1) edge node {$v_1 v_2$} (out);
		\end{tikzpicture}
		\subcaption{Copyless SST $S$.}
		\label{subfig:copyless}
	\end{subfigure}%
	\begin{subfigure}{0.5\linewidth}
		\centering
		\begin{tikzpicture}[every node/.style = {inner sep = 1pt, outer sep = 1pt, minimum size = 0pt, fill = white}]
			\node[state, accepting, initial] (1) {};
			\node (out) [right of = 1] {};

			\path[->] (1) edge[loop above] node {$a \begin{cases} v_1 \coloneq v_1 a v_2 \\ v_2 \coloneq v_2 a v_1 \end{cases}$} (1);
			\path[->] (1) edge node {$v_1 a a v_2$} (out);
		\end{tikzpicture}
		\subcaption{Copyful SST $T$.}
		\label{subfig:copyful}
	\end{subfigure}
\caption{
	Representative examples of copyless and copyful SSTs.
	In \eqref{subfig:copyless}, $\sigma$ denotes an arbitrary symbol of an arbitrary alphabet.
}
\label{fig:copying}
\end{figure}

An SST is \emph{copyless} if each variable is used at most once across all variable updates occurring on a single transition.
The copyless restriction prohibits updates involving assignments such as
\begin{itemize}
	\item $x \coloneq a y b y c$ and $y \coloneq x$, in which two copies of $y$ flow into $x$, and
	\item $x \coloneq y$ and $y \coloneq y$, where one copy of $y$ flows into each of the variables $x$ and $y$.
\end{itemize}
An SST that is not copyless is called \emph{copyful} \cite{FiliotReynier2017,FiliotReynier2021}.
\autoref{subfig:copyless} shows an example, $S$, of a copyless SST implementing the function $w \mapsto ww$.
\autoref{subfig:copyful} provides an example, $T$, of a copyful SST that implements the mapping $a^k \mapsto a^{2^{k+1}}$, which is not definable by any copyless SST.

The behavior of SSTs with respect to variable copying is particularly significant when it comes to characterizing the expressive power of the model.
Copyful SSTs, as evidenced by \autoref{fig:copying}, can model transductions having output strings of lengths exponential in the lengths of corresponding input strings.
Copyless SSTs, on the other hand, have outputs with lengths at most linear in the lengths of the corresponding inputs.
As shown by Alur, et al. \cite{AlurCerny2010,AlurDeshmukh2011}, the transductions definable by copyless SSTs coincide with those definable by monadic second-order logic graph transducers (MSOTs) \cite{Courcelle1992,Courcelle1994,CourcelleEngelfriet2012}.
The assumption that variable updates are copyless is crucial for this equivalence, because the linear upper bound on lengths of output strings is a fundamental property of MSOTs.

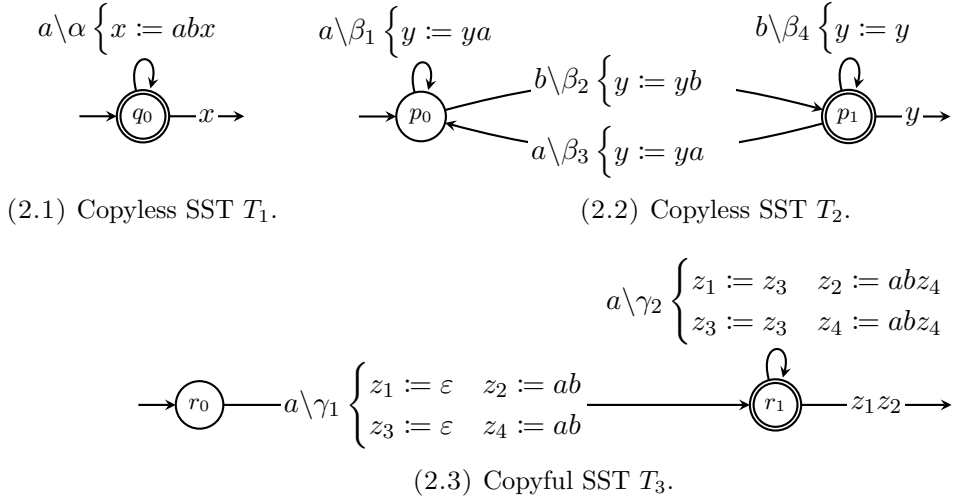
\begin{figure}
	\centering
	\begin{subfigure}{0.3\linewidth}
		\centering
		\begin{tikzpicture}[every node/.style = {inner sep = 1pt, outer sep = 1pt, minimum size = 0pt, fill = white}]
			\node[state, accepting, initial] (1) {$q_0$};
			\node (out) [right = 1cm of 1] {};

			\node (0)[below = 0.4cm of 1] {};

			\path[->] (1) edge[loop above] node {$a \backslash \alpha \begin{cases} x \coloneq abx \end{cases}$} (1);
			\path[->] (1) edge node {$x$} (out);
		\end{tikzpicture}
		\subcaption{Copyless SST $T_1$.}
		\label{subfig:T1}
	\end{subfigure}%
    \begin{subfigure}{0.7\linewidth}
		\begin{tikzpicture}[every node/.style = {inner sep = 1pt, outer sep = 1pt, minimum size = 0pt, fill = white}]
			\node[state, initial] (1)               {$p_0$};
			\node[state, accepting] (2)[right = 5cm of  1] {$p_1$};
			\node (out) [right = 1cm of 2] {};

			\path[->] (1) edge[loop above] node {$a \backslash \beta_1 \begin{cases} y \coloneq ya \end{cases}$} (1);
			\path[->] (1) edge[bend left = 15] node {$b \backslash \beta_2 \begin{cases} y \coloneq yb \end{cases}$}  (2);
			\path[->] (2) edge[bend left = 15] node {$a \backslash \beta_3 \begin{cases} y \coloneq ya \end{cases}$}  (1);
			\path[->] (2) edge[loop above] node {$b \backslash \beta_4 \begin{cases} y \coloneq y \end{cases}$} (2);
			\path[->] (2) edge node {$y$} (out);
		\end{tikzpicture}
		\subcaption{Copyless SST $T_2$.}
		\label{subfig:T2}
	\end{subfigure}

	\vspace{2ex}

	\begin{subfigure}{\linewidth}
		\centering
		\begin{tikzpicture}[every node/.style = {inner sep = 1pt, outer sep = 1pt, minimum size = 0pt, fill = white}]
			\node[state, initial]  (0) {$r_0$};
			\node[state, accepting] (1) [right = 7cm of 0] {$r_1$};
			\path[->] (0) edge node[pos = 0.4] {$a \backslash \gamma_1 \begin{cases}
					z_1 \coloneq \varepsilon & z_2 \coloneq ab \\
					z_3 \coloneq \varepsilon & z_4 \coloneq ab
				\end{cases}$} (1);
			\path[->] (1) edge[loop above] node {$a \backslash \gamma_2 \begin{cases}
				z_1 \coloneq z_3 & z_2 \coloneq abz_4 \\
				z_3 \coloneq z_3 & z_4 \coloneq abz_4
				\end{cases}$} (1);
			\node (out)[right = 2cm of 1] {};
			\path[->] (1) edge node {$z_1 z_2$} (out);
		\end{tikzpicture}
		\subcaption{Copyful SST $T_3$.}
		\label{subfig:T3}
	\end{subfigure}
	\caption{
		Two copyless SSTs $T_1$, $T_2$ and a copyless SST $T_3$ implementing their composition.
		An edge labeled by $a \backslash \alpha$ indicates that the machine transitions from the edge's source state to its target state upon reading symbol $a$ and applies the assignment $\alpha$ to its set of variables.
	}
	\label{fig:Alur_comp}
\end{figure}

As MSOTs are closed under composition\footnote{\cf{} Proposition 3.2 of \cite{Courcelle1992,Courcelle1994} and Theorem 7.14 of \cite{CourcelleEngelfriet2012}.}, it follows from their correspondence that the functions and relations definable by SSTs are also closed under composition.
Effective procedures are given in \cite{AlurCerny2010,AlurDeshmukh2011} to construct a copyless SST that realizes the composition of two copyless SSTs given as input.
Joost Englefriet observed\footnote{via a private correspondence with the authors of \cite{AlurDeshmukh2011}}, however, that these constructions can produce transducers with copyful assignments. 
Consider, for instance, the three deterministic SSTs shown in \autoref{fig:Alur_comp}, where $T_3$ is the composition of $T_1$ and $T_2$, constructed according to the methods provided by \cite{AlurCerny2010,AlurDeshmukh2011}.
The transducer $T_1$ implements the function $a^n \mapsto \paren{ab}^n$, while the transducer $T_2$ models the mapping $a^{n_1} b^{m_1} \dots a^{n_k} b^{m_k} \mapsto a^{n_1} b \dots a^{n_k} b$.
Notice that both $T_1$ and $T_2$ are copyless, yet there is a transition in $T_3$ with a copyful assignment $\gamma_2$.
On the other hand, notice how none of the variables copied by $\gamma_2$ are ever combined in any later assignment.

\begin{figure}
	\begin{subfigure}{0.5\linewidth}
		\centering
		\begin{tikzpicture}[every node/.style  = {font = \normalsize}]
			\node (1) {$v_1$};
			\node (2)[below = 1cm of 1] {$v_2$};
			\node (3)[right of = 1] {$v_1$};
			\node (4)[below = 1cm of 3] {$v_2$};
			\node (5)[right of = 3] {$v_1$};
			\node (6)[below = 1cm of 5] {$v_2$};
			
			\node (0)[below = 0.15cm of 4] {};

			\path[->] (1) edge (3);
			\path[->] (1) edge (4);
			\path[->] (2) edge (3);
			\path[->] (2) edge (4);
			\path[->] (3) edge (5);
			\path[->] (3) edge (6);
			\path[->] (4) edge (5);
			\path[->] (4) edge (6);
		\end{tikzpicture}
		\subcaption{Two-step flow graph in $T$.}
		\label{subfig:diamond}
	\end{subfigure}%
	\begin{subfigure}{0.5\linewidth}
		\centering
		\begin{tikzpicture}[node distance = 0.75cm, every node/.style  = {font = \normalsize}]
			\node (1) {$z_1$};
			\node (2)[below of = 1] {$z_2$};
			\node (3)[below of = 2] {$z_3$};
			\node (4)[below of = 3] {$z_4$};
			\node (5)[right = 2cm of 1] {$z_1$};
			\node (6)[below of = 5] {$z_2$};
			\node (7)[below of = 6] {$z_3$};
			\node (8)[below of = 7] {$z_4$};
			\node (9)[right = 2cm of 5] {$z_1$};
			\node (10)[below of = 9] {$z_2$};
			\node (11)[below of = 10] {$z_3$};
			\node (12)[below of = 11] {$z_4$};

			\path[->] (3) edge (5);
			\path[->] (3) edge (7);
			\path[->] (4) edge (6);
			\path[->] (4) edge (8);
			\path[->] (7) edge (9);
			\path[->] (7) edge (11);
			\path[->] (8) edge (10);
			\path[->] (8) edge (12);
		\end{tikzpicture}
		\subcaption{Flow graph for $\gamma_2 \circ \gamma_2$ in $T_3$.}
		\label{subfig:no_diamond}
	\end{subfigure}
	\caption{
		A pair of two-step flow graphs from $T$ and $T_3$.
		Four diamonds are present in \eqref{subfig:diamond}.
		Zero diamonds are present in \eqref{subfig:no_diamond}.
	}
	\label{fig:anti-confluence}
\end{figure}
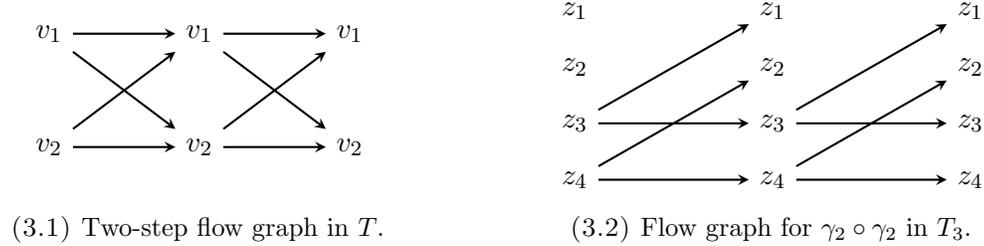

Directed acyclic graphs are a useful tool for representing the dynamics of variable copying in SSTs, facilitating both analysis and visualization.
The graphs depicted in \autoref{fig:anti-confluence}, for instance, provide convenient encodings of how information flows between variables of SSTs over the course of a computation.
In these illustrations, an edge $x \to y$ corresponds to a transition that involves an assignment to variable $y$ in which $x$ occurs.
\autoref{subfig:diamond} shows how variables flow into other variables during two consecutive applications of the single assignment of $T$, while \autoref{subfig:no_diamond} shows this for two consecutive applications of the assignment $\gamma_2$ of $T_3$.
Despite $\gamma_2$ being copyful, the diagram makes clear that there is no possibility of two copies of $z_3$ or $z_4$ converging and contributing to a common variable at a later point in time.
We call copyful SSTs exhibiting this type of behavior \emph{diamond-free}.
Intuitively, an SST is diamond-free if there is at most one path between any two vertices in the assignment graph induced by any computation.
The diamond-free restriction, while weaker than the copyless restriction, preserves its semantic intent: that outputs are linear in the lengths of their corresponding inputs.

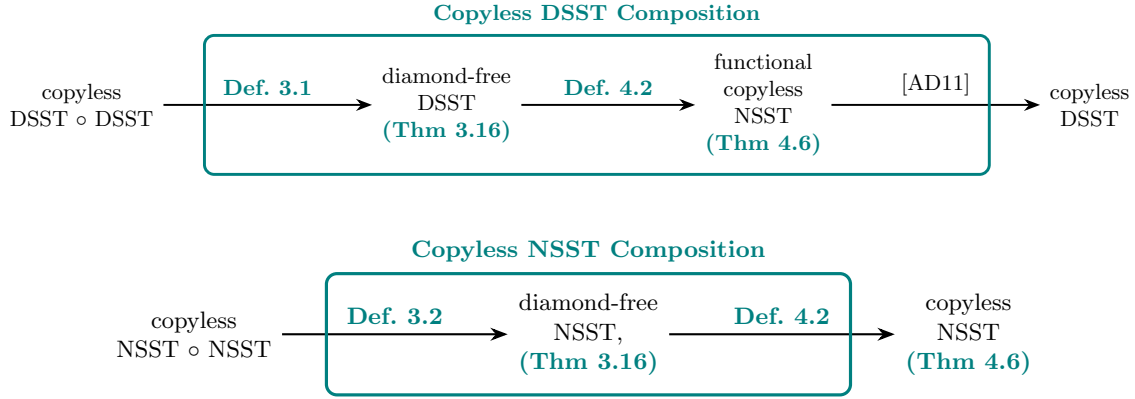
\begin{figure}
	\centering
	\resizebox{\linewidth}{!}{%
		\begin{tikzpicture}[> = Stealth]
			\node (di) [align=center] {copyless \\ DSST $\circ$ DSST};
			\node (ddf) [right = 3cm of di, align=center] {diamond-free \\ DSST \\ \teal{\bf (Thm~\ref{theorem:diamond-free})}};
			\node (dnfc) [right = 2.5cm of ddf, align=center] {functional \\ copyless \\ NSST \\ \teal{\bf (Thm~\ref{theorem:diamond-free_to_copyless})}};
			\node (dc) [right = 3cm of dnfc, align=center] {copyless \\ DSST};

			\path[->] (di) edge node [above, name=DC, teal] {\bf Def.~\ref{def:dsst_comp}} (ddf);
			\path[->] (ddf) edge node [above, name=DDFC, teal] {\bf Def.~\ref{def:diamond-free_to_copyless}} (dnfc);
			\path[->] (dnfc) edge node[above, name=AD11a]{\cite{AlurDeshmukh2011}} (dc);

			\node[teal, very thick, rounded corners, fit=(DC)(DDFC)(ddf)(dnfc)(AD11a), draw, label=\teal{\textbf{Copyless DSST Composition}}] {};

			\node[below = 1.5cm of DDFC] {};
		\end{tikzpicture}
	}
	
	\centering
	\begin{tikzpicture}[> = Stealth]
		\node (ni) [align=center] {copyless \\ NSST $\circ$ NSST};
		\node (ndf) [right = 3cm of ni, align=center] {diamond-free \\ NSST, \\ \teal{\bf (Thm~\ref{theorem:diamond-free})}};
		\node (nc) [right = 3cm of ndf, align=center] {copyless \\ NSST \\ \teal{\bf (Thm~\ref{theorem:diamond-free_to_copyless})}};

		\path[->] (ni) edge node [above, name=NC, teal] {\bf Def.~\ref{def:nsst_comp}} (ndf);
		\path[->] (ndf) edge node [above, name=NDFC, teal] {\bf Def.~\ref{def:diamond-free_to_copyless}} (nc);

		\node[teal, very thick, rounded corners, fit=(ndf)(NC)(NDFC), draw, label=\teal{\textbf{Copyless NSST Composition}}] {};
	\end{tikzpicture}
	\caption{Schematic of this paper's approach to composing copyless SSTs.}
	\label{fig:complete-comp}
\end{figure}

\paragraph{Contributions and Outline.}
In \autoref{sec:sst}, we introduce the basic definitions and ideas around SSTs.
In \autoref{sec:composition}, we present a slightly modified and more detailed account of the composition constructions of Alur, et al. for both DSSTs and NSSTs.
We prove their correctness and establish bounds on the size of the resulting transducers in \autoref{subsec:analysis}.
In \autoref{subsec:copyless}, show that the composite transducers are necessarily diamond-free. 
Lastly, in \autoref{sec:toFNSST}, we develop a construction that transforms any diamond-free NSST into an equivalent copyless NSST. 
As a result, the removal of noncopyless assignments can be added as post-processing step, thereby yielding a complete direct method for SST composition.
A diagrammatic summary of our results is given in \autoref{fig:complete-comp}.
By recovering a composition construction that acts directly on copyless SSTs, we solidify the foundations of works depending on this operation, such as \cite{ThakkarKanadeAlur2013,DaveDohmenKrishnaTrivedi2021}.

\paragraph{Related Work.}
The diamond-free property has been considered in related literature \cite{LopezMonmegeTalbot2019} studying determinization of finitely-ambiguous cost register automata.
While our notion of diamond-freeness essentially coincides with the notion considered in that work, our results are independent and apply in a distinct context.

An alternative restriction on copying behavior in SSTs, called \emph{bounded-copy}, is also considered in related work \cite{AlurFiliotTrivedi2012}.
An SST is bounded-copy, if in the assignment graph for any computation, there are finitely many paths between any two vertices.
For any non-negative integer $k$, a bounded copy SST is called $k$-copy if there are at most $k$ paths between any two vertices in the assignment graph induced by a computation.
From this definition, it is clear that copyless SSTs correspond to 0-copy SSTs and diamond-free SSTs correspond to 1-copy SSTs.
Note, however, that \cite{AlurFiliotTrivedi2012} considers deterministic SSTs while the present work considers nondeterministic SSTs.
Furthermore, the methods we use to convert diamond-free NSSTs into copyless NSSTs are simpler and distinct from those employed in \cite{AlurFiliotTrivedi2012} to convert bounded-copy DSSTs into copyless DSSTs.

\section{Streaming String Transducers}
\label{sec:sst}
The natural numbers will be written as $\bb{N}$.
The cardinality of a set $X$ is denoted by $\card{X}$, and its powerset by $2^X$.

Let $\Sigma^*$ be the set of all strings over a finite alphabet $\Sigma$ and let $\varepsilon$ be the empty string.
For $s \in \Sigma$ and $w \in \Sigma^*$, let $s \in w$ indicate that the symbol $s$ occurs in the string $w$, let $|w|$ denote the length of $w$, and let $|w|_s$ be the number of occurrences of the symbol $s$ within $w$.
We use product notation to denote iterated concatenation, so that $\prod^n_{k=1} w_k = w_1 w_2 \dots w_n$.

A morphism, over alphabets $\Sigma$ and $\Lambda$, is a function $h : \Sigma^* \to \Lambda^*$ which respects concatenation such that $h(\varepsilon) = \varepsilon$ and $h(w_1w_2) = h(w_1)h(w_2)$, for any $w_1, w_2 \in \Sigma^*$.
Every morphism is completely specified by its restriction to $\Sigma$.
For a set $X$, let $\ms{Id}_X : X \to X$ be the identity function that maps every element $x$ to itself.
With respect to a subset $X \subseteq \Sigma$, let $\ms{Er}_X : \Sigma^* \to \paren{\Sigma \smallsetminus X}^*$ be the erasing morphism, defined such that $\ms{Er}_X(x) = \varepsilon$ if $x \in X$ and $\ms{Er}_X(s) = s$ if $s \in \Sigma \smallsetminus X$.

For a function $f : X \to Y$, we write $\dom{f}$ for its domain and $\im{f}$ for its image.
The composition $g \circ f: X \to Z$ of two functions $f: X \to Y$ and $g: Y \to Z$ is defined in the standard fashion as $x \mapsto g(f(x))$.
We adopt the same notation to refer to domains, images, and compositions of relations.

\begin{defi}[SST Syntax]
	A streaming string transducer $T$ is given by a tuple $\tup{\Sigma, \Lambda, Q, q_0, X, A, \Delta, F}$ where $\Sigma$ is a finite input alphabet, $\Lambda$ is a finite output alphabet, $Q$ is a finite set of states, $q_0 \in Q$ is a distinguished initial state, $X$ is a finite set of string variables, $A$ is a finite set of assignment maps with type $X \to (\Lambda \cup X)^*$ for updating variables, $\Delta \subseteq Q \times \Sigma \times A \times Q$ is a transition relation, and $F : Q \to (\Lambda \cup X)^*$ is a partial output function.
\end{defi}

An assignment $\alpha$ of type $X \to (\Lambda \cup X)^*$ can naturally be extended to a morphism $(\Lambda \cup X)^*  \to (\Lambda \cup X)^*$ such that $\alpha(s) = s$, for $s \in \Lambda$ and $\alpha\paren{\prod^n_{k=1} w_k} = \prod^n_{k=1} \alpha(w_k)$, for words in $(\Lambda \cup X)^*$.
When viewed in this manner, assignments can be composed in sequence, and we set $\comp^n_{k=1} \alpha_k = \alpha_n \circ \alpha_{n-1} \circ \dots \circ \alpha_1$ as the notation for iterated composition.

The \emph{shape} of a string $w_0 \prod^n_{k=1} x_k w_k \in \paren{X \cup \Lambda}^*$ is a summary $\square x_1 \square x_2 \dots \square x_n \square$ of the string's topology, which ignores details of the variable-free substrings occurring between variables.
Here, the symbol $\square$ simply indicates the presence of a substring $w_k \in \Lambda^*$.
Such shapes are entirely determined by the order of the variables occurring in the original string.

A run of an SST on an input word $w = w_1 \dots w_n \in \Sigma^*$ is a finite sequence of transitions $q_0 \xrightarrow[\alpha_1]{w_1} \dots \xrightarrow[\alpha_n]{w_n} q_{n}$ such that $\tup{q_{k-1}, w_k, \alpha_k, q_k} \in \Delta$ for all $1 \leq k \leq n$.
For a run $\rho$, define the valuation $\mc{V}_\rho : \paren{X \cup \Lambda}^* \to \Lambda^*$ as $\comp^n_{k=1} \alpha_k$.

\begin{defi}[SST Semantics]
	An SST $T$ specifies a relation $\sem{T} \subseteq \Sigma^* \times \Lambda^*$ where
	\begin{equation*}
		\sem{T} = \set{\tup{w, \ms{Er}_X \circ \mc{V}_\rho \circ F(q_n)} : \tn{exists run } \rho \tn{ in } T \tn{ on } w \tn{ ending in } q_n \in \dom{F}}.
	\end{equation*}
\end{defi}

An SST is deterministic (DSST), if for every pair $\tup{q, s} \in Q \times \Sigma$ there is exactly one transition $\tup{q, s, \alpha, q'} \in \Delta$, and is nondeterministic (NSST) otherwise.
A \emph{functional} NSST is one for which every element of $\dom{\sem{T}}$ maps to a unique element of $\im{\sem{T}}$. 

\subsection{Copying Variables}

A string $w \in (\Sigma_1 \cup \Sigma_2)^*$ is copyless in $\Sigma_1$ if each element of $\Sigma_1$ occurs at most once in $w$, \ie{} when $\max_{s \in \Sigma_1} \card{w}_s \leq 1$.
We write $\brak{\Sigma_1}$ for the set of copyless strings over $\Sigma_1$ and $\brak{\Sigma_1,\Sigma_2}$ for set of strings over $(\Sigma_1 \cup \Sigma_2)^*$ that are copyless in $\Sigma_1$.

An assignment $\alpha \in A$ is copyless if the string $\prod^{\card{X}}_{k=1} \alpha(x_k)$ is copyless in $X$, making it an element of $\brak{X, \Lambda}$.
It is straightforward to observe that any composition of copyless assignments is also copyless.
We say an SST is copyless if, for all transitions $\tup{q, s, \alpha, q'} \in \Delta$, the assignment $\alpha$ is copyless.
An SST is copyful if it has at least one copyful assignment.

The following proposition characterizes the cardinality of the set of all copyless strings over a finite alphabet in terms of the size of that alphabet.
It will be used to establish bounds on the size of composite SSTs in \autoref{subsec:analysis}.

\begin{prop}
	\label{proposition:copyless_card_bound}
	If $\card{X} = n$ and $n \geq 2$, then $\card{\brak{X}} = \floor*{e n!}$.
\end{prop}
\begin{proof}
	Strings in $\brak{X}$ can be any length less than or equal to $n$, so counting the number of elements amounts to counting the number of ways to choose---without repetition---and order $k$ elements from $X$ for $0 \leq k \leq n$.
	Thus,
	\begin{equation*}
		\card{\brak{X}} \quad=\quad \sum^{n}_{k=0} \frac{n!}{(n-k)!} \quad=\quad n! \sum^{n}_{k=0} \frac{1}{(n-k)!} \quad=\quad n! \sum^{n}_{k=0} \frac{1}{k!}.
	\end{equation*}
	By Taylor's theorem \cite{Rudin1976}, $e = \sum^{n}_{k=0} \frac{1}{k!} + \frac{e^a}{(n+1)!}$, for some $a \in (0,1)$, and so we obtain the equation
	\begin{equation*}
		\card{\brak{X}} \quad=\quad n! \left(e - \frac{e^a}{(n+1)!}\right) \quad=\quad e n! - \frac{e^a}{n+1}.
	\end{equation*}
	Since the map $a \mapsto e^a$ is increasing on the unit interval, the subtrahend of the above difference may be bounded as
	\begin{equation*}
		\frac{1}{n+1} \quad \leq \quad \frac{e^a}{n+1} \quad \leq \quad \frac{e}{n+1}.
	\end{equation*}
	Furthermore, the assumption that $n \geq 2$ implies that $\frac{e^a}{n+1} < 1$.
	Finally, $\brak{X}$ is a finite set and thus has cardinality in the positive integers, so we conclude that $\card{\brak{X}} = \floor*{e n!}$.
\end{proof}

\subsection{Flow Graphs and Diamonds}

The copying behavior of an SST is characterized by \emph{flow graphs} associated to its assignments.

\begin{defi}[Flow Graph]
    Let $\alpha = \alpha_1 \dots \alpha_n$ be a word over $A^*$. 
    The flow graph of $\alpha$ is a directed acyclic graph $G(\alpha) = \tup{V(\alpha), E(\alpha)}$ such that 
    \begin{itemize}
        \item the set of vertices $V(\alpha) = \biguplus_{k=1}^{n+1} X_k$ is made up of $n+1$ disjoint copies of $X$ and
        \item $\tup{x,y} \in E(\alpha)$ if there exists $1 \leq k \leq n$ such that $x \in X_k$, $y \in X_{k+1}$, and $x \in \alpha_k(y)$.
    \end{itemize}
\end{defi}

The flow graph $G(\rho)$ of a run $\rho = q_0 \xrightarrow[\alpha_1]{w_1} \dots \xrightarrow[\alpha_n]{w_n} q_{n}$ is identified with the flow graph $G(\alpha_1 \dots \alpha_n)$ of its sequence of assignments.

\begin{defi}[Diamond]
    Let $\alpha = \alpha_1 \dots \alpha_n$ be a word over $A^*$, and let $G(\alpha)$ be its flow graph.
    A diamond is a subgraph of $G(\alpha)$, consisting of two distinct paths from a source vertex $s \in X_i$ to a target vertex $t \in X_j$, with $i < j$.
\end{defi}

\paragraph{Diamond-Freeness.}
An assignment $\alpha$ is diamond-free if $\card{\alpha(x)}_y \leq 1$ holds for all $x,y \in X$ or, equivalently, if $G(\alpha)$ has no double edges.
Copyless assignments are diamond-free, but diamond-free assignments are not necessarily copyless.
A run $\rho$ is diamond-free if there is at most one path between any two vertices in $G(\rho)$.
An SST is diamond-free if all of its runs are diamond-free.
Note that runs and SSTs may fail to be diamond-free even if all assignments in $A$ are diamond-free.

\section{Composition Construction}
\label{sec:composition}
This section presents a modified and elaborated treatment of the composition constructions of Alur, et al. \cite{AlurCerny2010,AlurDeshmukh2011}. 
Our departure from the original versions is centered around two aspects:
\begin{enumerate}
	\item we impose a strict indexing of the variables of the composite SST that them to be partitioned based on their roles in the process, and
	\item we provide explicit detailed definitions, in the form of recursive higher-order functions, of the operators used to produce the composite SST.
\end{enumerate}
The indexing scheme in combination with the recursively defined operators allow us to establish the diamond-free property in an almost type-directed manner.

We begin by presenting the construction under the assumption that the given SSTs are deterministic in \autoref{subsec:dsst_comp}.
Then, in \autoref{subsec:nsst_comp}, we describe the necessary adaptations to the procedure for NSSTs.
In \autoref{subsec:analysis}, we prove correctness of the modified constructions and establish bounds on the size of the composite SSTs.
Lastly, in \autoref{subsec:copyless}, we prove that the composite transducers must be diamond-free.

\subsection{DSST Composition}
\label{subsec:dsst_comp}

Fix copyless DSSTs $T^1_2 = \tup{\Sigma_1, \Sigma_2, Q, q_0, X, A, \Delta^1_2, F^1_2}$ and $T^2_3 = \tup{\Sigma_2, \Sigma_3, P, p_0, Y, B, \Delta^2_3, F^2_3}$ as inputs.
The DSST $T^1_3 = \tup{\Sigma_1, \Sigma_3, R, r_0, \mc{Z}, C, \Delta^1_3, F^1_3}$ will be the composite transducer.
The set of variables of $T^1_3$ is written as $\mc{Z}$ to allow the use of uppercase $Z$ for individual variables, which makes our index notation more readable.

The main idea underlying the composition constructions is to simulate the state transitions of $T^1_2$ while keeping a record of what $T^2_3$ does when reading each possible variable in $X$ from each possible state in $P$. 
Since $T^1_2$ builds its output piecewise, these records are critical.  
By storing the actions of $T^2_3$ for each possible eventuality, it is possible to then chain together those actions which correspond to the run of $T^2_3$ on the output of $T^1_2$.
In particular, the composite transducer incorporates two types of \emph{summaries} as finite maps in the states of $T^1_3$.
\begin{enumerate}
	\item \emph{State summaries} are maps $f : (P \times X) \to P$ that record the state in $T^2_3$ that is reached after reading the contents of a variable $x$, for every possible combination of starting states in $p \in P$ of $T^2_3$ and variables $x \in X$ of $T^1_2$.
	\item \emph{Shape summaries} are maps $g: (P \times X \times Y) \to \brak{Y \cup \mc{Z}}$ that store the shape of each variable $y \in Y$ after reading a variable $x \in X$ from a starting point $p \in P$, for every possible combination of parameters. 
\end{enumerate}
Since their domains and ranges are finite, each summary is finitely representable and there are finitely many summaries of each type. 
We write $\bb{F}$ for the set of all functions with type $(P \times X) \to P$ and $\bb{G}$ for the set of all functions of type $(P \times X \times Y) \to \brak{Y \cup \mc{Z}}$.
The variables $\mc{Z} = \set{Z^{p,x}_{y,0}, Z^{p,x}_{y,1} : \tup{p,x,y} \in P \times X \times Y}$ of $T^1_3$ are used by the shape summary maps to store variable-free substrings of shapes.

We now define the higher-order operators that will be used to explicitly define the composition constructions.
For the remainder of this section, we set the following conventions.
\begin{itemize}
	\item $\mb{x}$ represents strings from $\brak{X, \Sigma_2}$.
	\item $\mb{y}$ denotes strings from $\brak{Y \cup \mc{Z},\Sigma_3}$ and $\brak{Y,\Sigma_3}$.
	\item $\mb{z}$ represents strings from $\brak{\mc{Z}, \Sigma_3}$.
	\item $g^p_x$ is shorthand for the map $y \mapsto g(p,x,y)$, where $g \in \bb{G}$.
\end{itemize}

\paragraph{State Transition Summarizer.} 
The map $\mc{F} : P \times \brak{X, \Sigma_2} \times \bb{F} \to P$ is responsible for updating destination states in $T^2_3$ based on updated variables in $T^1_2$ and is defined as
\begin{equation}
\label{eq:fhat_dsst}
	\mc{F}(p, \mb{x}, f) =
	\begin{cases}
		p &\tn{if } \mb{x} = \varepsilon  \\[1ex]
		\mc{F}\paren{p', \mb{x}', f} &\tn{if }  \mb{x} = s \: \mb{x}' \tn{ and } \tup{p, s, \alpha, p'} \in \Delta^2_3 \\[1ex]
		\mc{F}\paren{f(p, x), \mb{x}', f} &\tn{if } \mb{x} = x \: \mb{x}'.
	\end{cases}
\end{equation}

\paragraph{Assignment Summarizer.} 
The processes that create shape summaries and assignments share a dependency on the map $\mc{G} : P \times \brak{X, \Sigma_2} \times \bb{F}  \times \bb{G} \to \paren{Y \to \brak{Y \cup \mc{Z},\Sigma_3}}$ defined as
\begin{equation}
\label{eq:ghat_dsst}
	\mc{G}(p, \mb{x}, f, g) =
	\begin{cases}
		\ms{Id}_Y & \text{if } \mb{x} = \varepsilon  \\[1ex]
		\beta \circ \mc{G}\paren{p', \mb{x}', f, g}  &\tn{if } \mb{x} = s \: \mb{x}' \tn{ and } \tup{p, s, \beta, p'} \in \Delta^2_3 \\[1ex]
		g^p_x \circ \mc{G}\paren{f(p,x), \mb{x}', f, g} &\tn{if } \mb{x} = x \: \mb{x}'.
	\end{cases}
\end{equation}

\paragraph{Shape Generator.}
The map $\mc{S} : P \times X \times \paren{Y \to \brak{Y \cup \mc{Z},\Sigma_3}} \to \paren{Y \to \brak{Y \cup \mc{Z}}}$ computes shape summaries and is given as
\begin{equation}
	\label{eq:shape}
	\mc{S}(p, x, h) = y \mapsto \paren{\prod^n_{k=1} Z^{p, x}_{y_k,0} \: y_k} Z^{p, x}_{y,1} \qquad \tn{ if } h(y) = \mb{z}_0 \prod^n_{k=1} y_k \: \mb{z}_k.
\end{equation}

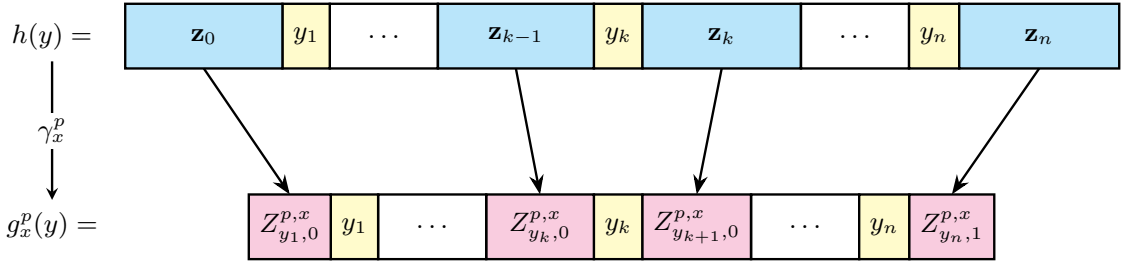
\begin{figure}
	\centering
	\resizebox{\linewidth}{!}{
		\begin{tikzpicture}[block/.style={minimum height=2em,outer sep=0pt,draw,rectangle,node distance=0pt}]
			\node[block, fill = yellow!25] (A) {$y_k$};
			\node[block, fill = cyan!25] (B) [left=of A] {$\quad\:\: \mb{z}_{k-1} \quad\:\:$};
			\node[block] (C) [left=of B] {$\quad \dots \quad$};
			\node[block, fill = cyan!25] (D) [right=of A] {$\qquad \mb{z}_k \qquad$};
			\node[block] (E) [right=of D] {$\quad \dots \quad$};
			\node[block, fill = yellow!25] (F) [left=of C] {$y_1$};
			\node[block, fill = cyan!25] (G) [left=of F] {$\qquad \mb{z}_0 \qquad$};
			\node[block, fill = yellow!25] (H) [right=of E] {$y_n$};
			\node[block, fill = cyan!25] (I) [right=of H] {$\qquad \mb{z}_n \qquad$};

			\node[block, fill = yellow!25] (AA) [below = 1.5cm of A] {$y_k$};
			\node[block, fill = magenta!25] (BB) [left=of AA] {$\:\: Z^{p,x}_{y_k, 0} \:\:$};
			\node[block] (CC) [left=of BB] {$\quad \dots \quad$};
			\node[block, fill = magenta!25] (DD) [right=of AA] {$Z^{p,x}_{y_{k+1}, 0}$};
			\node[block] (EE) [right=of DD] {$\quad \dots \quad$};
			\node[block, fill = yellow!25] (FF) [left=of CC] {$y_1$};
			\node[block, fill = magenta!25] (GG) [left=of FF] {$Z^{p,x}_{y_1, 0}$};
			\node[block, fill = yellow!25] (HH) [right=of EE] {$y_n$};
			\node[block, fill = magenta!25] (II) [right=of HH] {$Z^{p,x}_{y_n, 1}$};

			\node (g) [left = 0.25cm of G] {$h(y) =$};
			\node (sh) [below = 1.65cm of g] {$g^p_x(y) =$};

			\path[->] (g) edge node[inner sep = 2pt, outer sep = 2pt, minimum size = 0pt, fill = white] {$\gamma^p_x$} (sh);

			\path[->, > = Stealth] (G.south) edge (GG.north);
			\path[->, > = Stealth] (B.south) edge (BB.north);
			\path[->, > = Stealth] (D.south) edge (DD.north);
			\path[->, > = Stealth] (I.south) edge (II.north);



			
		\end{tikzpicture}
	}
	\caption{An illustration of the relationship between an assignment summary $h = \mc{G}\paren{p, \alpha(x), f, g}$ and the corresponding shape $g^p_x = \mc{S}^p_x(h)$ and assignment $\gamma^p_x = \mc{A}^p_x(h)$.}
	\label{fig:trifit}
\end{figure}

\paragraph{Assignment Generator.}
The map $\mc{A} : P \times X \times \paren{Y \to \brak{Y \cup \mc{Z},\Sigma_3}} \to \paren{\mc{Z} \to \brak{\mc{Z}, \Sigma_3}}$ computes assignments and is given as
\begin{equation}
	\label{eq:assign}
	\mc{A}(p,x,h) = 
	\begin{cases}
		Z^{p,x}_{y,0} \mapsto \varepsilon &\tn{if } y \notin h(y'), \tn{ for all } y' \in Y \\[1ex]
		Z^{p,x}_{y,0} \mapsto \mb{z} &\tn{if } h(y') = \mb{z} \: y \: \mb{y} \tn{ or } y'' \: \mb{z} \: y \in h(y') \tn{ for some } y', y'' \in Y \\[1ex]
		Z^{p,x}_{y,1} \mapsto \mb{z} &\tn{if } h(y) = \mb{z} \tn{ or } h(y) = \mb{y} \: y' \: \mb{z}, \tn{ for some } y' \in Y. 
	\end{cases}
\end{equation}
We write $\mc{S}^p_x$ and $\mc{A}^p_x$, respectively, for the mappings $h \mapsto \mc{S}(p,x,h)$ and $h \mapsto \mc{A}(p,x,h)$.
Letting
\begin{equation*}
	g^p_x = \mc{S}^p_x \circ \mc{G}\paren{p, \alpha(x), f, g} \quad\tn{ and }\quad \gamma^p_x = \mc{A}^p_x \circ \mc{G}\paren{p, \alpha(x), f, g},
\end{equation*}
the next equation holds for all $y \in Y$:
\begin{equation}
	\label{eq:AS_combine}
	\mc{G}\paren{p, \alpha(x), f, g}(y) = \gamma^p_x \circ g^p_x(y).
\end{equation}
Thus, the following diagram commutes, where $\pi_k$ and $\iota_k$ are the $k^\tn{th}$ canonical projection and injection.
This relationship is further illustrated in \autoref{fig:trifit}.
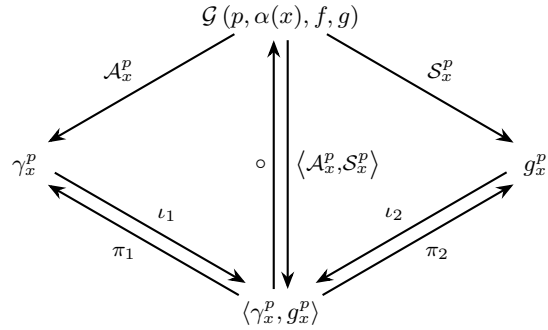
\begin{center}
	\begin{tikzcd}
		& & \mc{G}\paren{p, \alpha(x), f, g} \arrow[ddrr, thick, shift left, "{\mc{S}^p_x}"] \arrow[dddd, thick, shift left, "{\tup{\mc{A}^p_x, \mc{S}^p_x}}"] \arrow[ddll, thick, shift right, "{\mc{A}^p_x}"'] & & \\
		& & & & \\
		\gamma^p_x \arrow[ddrr, thick, shift left, "\iota_1"] & &  & & g^p_x \arrow[ddll, thick, shift right, "\iota_2"'] \\
		& & & & \\
		& & \tup{\gamma^p_x, g^p_x} \arrow[uurr, thick, shift right, "\pi_2"'] \arrow[uull, thick, shift left, "\pi_1"] \arrow[uuuu, thick, shift left, "\circ"] & &
	\end{tikzcd}
\end{center}

The following definition formalizes the construction of $T_3$ from DSSTs $T_1$ and $T_2$.

\begin{defi}[DSST Composition]
\label{def:dsst_comp}
	Suppose that $T^1_2 = \tup{\Sigma_1, \Sigma_2, Q, q_0, X, A, \Delta^1_2, F^1_2}$ and $T^2_3 = \tup{\Sigma_2, \Sigma_3, P, p_0, Y, B, \Delta^2_3, F^2_3}$ are an arbitrary pair of copyless DSSTs.
	Their composition is a DSST $T^1_3 = \tup{\Sigma_1, \Sigma_3, R, r_0, \mc{Z}, C, \Delta^1_3, F^1_3}$ where
	\begin{itemize}
		\item $R = Q \times \bb{F} \times \bb{G}$ such that $r_0 = \tup{q_0, f_0, g_0}$ with $f_0(p, x) = p$ and $g_0(p, x, y) = y$,
		\item $\mc{Z} = \set{Z^{p,x}_{y,0}, Z^{p,x}_{y,1} : \tup{p, x, y} \in P \times X \times Y}$,
		\item $C = \set{\gamma : \tup{r, s, \gamma, r'} \in \Delta^1_3}$,
		\item $\tup{\tup{q, f, g}, s, \gamma, \tup{q', f', g'}} \in \Delta^1_3$ if, there exists $\tup{q, s, \alpha, q'} \in \Delta^1_2$ such that
		\begin{enumerate}
			\item $f'(p,x) = \mc{F}(p, \alpha(x),f)$,
			\item $g'(p,x,y) = \mc{S}^p_x\paren{\mc{G}\paren{p, \alpha(x), f, g}}(y)$, and
			\item $\gamma\paren{Z^{p,x}_{y,k}} = \mc{A}^p_x\paren{\mc{G}\paren{p, \alpha(x), f, g}}\paren{Z^{p,x}_{y,k}}$,
		\end{enumerate}
		\item $F^1_3(q, f, g) = \ms{Er}_Y \circ \mc{G}\paren{p_0, F^1_2(q), f, g} \circ F^2_3 \circ \mc{F}\paren{p_0, F^1_2(q), f}$.
	\end{itemize}
\end{defi}

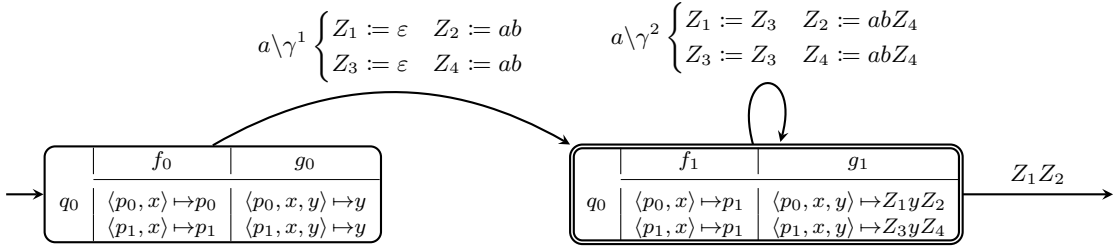
\begin{figure}
	\centering
	\begin{tikzpicture}
		\tikzset{
			state/.style={rectangle,rounded corners,draw=black,minimum height=2em, inner sep=0.5pt, text centered}
		}
		\node[state, initial, font = \scriptsize] (0) {
			\begin{tabular}{c | c | c}
				&  $f_0$ & $g_0$ \\
				\cmidrule{2-3}
				$q_0$ & $\tup{p_0, x} {\mapsto} p_0$ & $\tup{p_0, x, y} {\mapsto} y$ \\
				&$\tup{p_1, x} {\mapsto} p_1$ & $\tup{p_1, x, y} {\mapsto} y$
			\end{tabular}};
		\node[state, accepting, font = \scriptsize] (1)[right = 2.5cm of 0] {
			\begin{tabular}{c | c | c}
				& $f_1$ & $g_1$ \\
				\cmidrule{2-3}
				$q_0$ & $\tup{p_0, x} {\mapsto} p_1$ & $\tup{p_0, x, y} {\mapsto} Z_1yZ_2$ \\
				&$\tup{p_1, x} {\mapsto} p_1$ & $\tup{p_1, x, y} {\mapsto} Z_3yZ_4$
			\end{tabular}};
		\node (out)[right = 2cm of 1] {};

		\path[->] (0.north) edge[bend left] node[above] {$a \backslash \gamma^1 \begin{cases}
			Z_1 \coloneq \varepsilon & Z_2 \coloneq ab \\
			Z_3 \coloneq \varepsilon & Z_4 \coloneq ab 
			\end{cases}$} (1.north west);
		\path[->] (1) edge[loop above] node[above] {$a \backslash \gamma^2 \begin{cases}
			Z_1 \coloneq Z_3 & Z_2 \coloneq abZ_4 \\
			Z_3 \coloneq Z_3 & Z_4 \coloneq abZ_4 
			\end{cases}$} (1);
		\path[->] (1) edge node[above] {$Z_1 Z_2$} (out);
	\end{tikzpicture}
	\caption{
		A detailed view of internal maps contained in the states of $T_3$ from \autoref{fig:Alur_comp}, when constructed according to \autoref{def:dsst_comp}.
		Note that we omit unreachable states.
	}
\end{figure}

\subsection{NSST Composition}
\label{subsec:nsst_comp}

\begin{figure}
	\centering
	\begin{tikzpicture}
		\node[state,initial above] (1) {$p_1$};
		\node[state, accepting] (2)[left = 2cm of 1] {$p_2$};
		\node[state, accepting] (4)[above left = 0.05cm and 2cm of 2] {$p_4$};
		\node[state, accepting] (5)[below left = 0.05cm and 2cm of 2] {$p_5$};
		\node[state, accepting] (3)[right = 2cm of 1] {$p_3$};
		\node[state, accepting] (6)[above right = 0.05cm and 2cm of 3] {$p_6$};
		\node[state, accepting] (7)[below right = 0.05cm and 2cm of 3] {$p_7$};

		\path[->] (1) edge node[above] {$a \backslash \beta_1$} (2);
		\path[->] (1) edge node[above] {$a \backslash \beta_2$} (3);
		\path[->] (2) edge node[sloped,above] {$a \backslash \beta_3$} (4);
		\path[->] (2) edge node[sloped,below] {$a \backslash \beta_4$} (5);
		\path[->] (3) edge node[sloped,above] {$a \backslash \beta_5$} (6);
		\path[->] (3) edge node[sloped,below] {$a \backslash \beta_6$} (7);
	\end{tikzpicture}
	\caption{An NSST illustrating the need for $\mc{H}$.}
	\label{fig:impossible_pairs}
\end{figure}
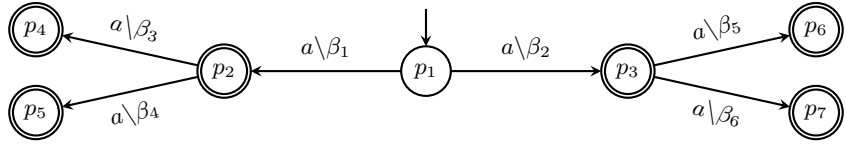

\autoref{def:dsst_comp} requires some adaptation for NSSTs.
One approach to this adaptation might be to generalize the maps $\mc{F}$ and $\mc{G}$ independently to accommodate nondeterminism.
This would involve defining variations of $\mc{F}$ and $\mc{G}$ having codomains $2^P$ and $2^{Y \to \brak{Y \cup \mc{Z}, \Sigma_3}}$, respectively.
Then, one could add transitions in $T^1_3$ whenever the summaries in the destination state could be obtained via these maps from the summaries in the source state.
Doing this, however, could lead to grouping a state summary based on one nondeterministic run in $T^2_3$ together with a shape summary based on a different nondeterministic run.
To see this, imagine that the NSST in \autoref{fig:impossible_pairs} is $T^2_3$.
Notice that the string $aa$ leads to four distinct states and induces four distinct assignments, but to faithfully simulate this transducer, each destination state should be paired to a unique assignment.
For instance, if the state summary points to $p_4$, then the only shape summary that correctly pairs with this state summary represents the effects of applying the assignment $\beta_3 \circ \beta_1$.
Proceeding as we have just described would lead to 3 additional incorrect pairings, and this would result in an incorrect composite transducer.

To avoid such situations, it is necessary to synchronize the computation of state summaries with the computation of shape summaries and assignments.
This synchronization is achieved by the map $\mc{H} : (P \times \brak{X, \Sigma_2} \times \bb{F} \times \bb{G} ) \to 2^{P \times (Y \to \brak{Y \cup \mc{Z}, \Sigma_3})}$ which combines $\mc{F}$ and $\mc{G}$ in more structured way and is defined as
\begin{equation*}
	\mc{H}(p, \mb{x}, f, g) =
	\begin{cases}
		\set{\tup{p, \ms{Id}_Y}} &\tn{if } \mb{x} = \varepsilon, \\[1ex]
		\bigcup\limits_{\tup{p, s, \alpha, p'} \in \Delta^2_3} \set{\tup{p'', \alpha \circ h} : \tup{p'', h} \in \mc{H}\paren{p', \mb{x}', f, g}} &\tn{if } \mb{x} = s \: \mb{x}', \\[3ex]
		\set{\tup{p', g^p_x \circ h} : \tup{p', h} \in \mc{H}\paren{f(p,x), \mb{x}', f, g}} &\tn{if } \mb{x} = x \: \mb{x}'.
	\end{cases}
\end{equation*}

\begin{defi}[NSST Composition]
\label{def:nsst_comp}
	Suppose that $T^1_2 = \tup{\Sigma_1, \Sigma_2, Q, q_0, X, A, \Delta^1_2, F^1_2}$ and $T^2_3 = \tup{\Sigma_2, \Sigma_3, P, p_0, Y, B, \Delta^2_3, F^2_3}$ are an arbitrary pair of copyless NSSTs.
	Their composition is an NSST $T^1_3 = \tup{\Sigma_1, \Sigma_3, R, r_0, \mc{Z}, C, \Delta^1_3, F^1_3}$ where
	\begin{itemize}
		\item $R = Q \times \bb{F} \times \bb{G}$ such that $r_0 = \tup{q_0, f_0, g_0}$ with $f_0(p, x) = p$ and $g_0(p, x, y) = y$,
		\item $\mc{Z} = \set{Z^{p,x}_{y,0}, Z^{p,x}_{y,1} : \tup{p, x, y} \in P \times X \times Y}$, 
		\item $C = \set{\gamma : \tup{r, s, \gamma, r'} \in \Delta^1_3}$,
		\item $\tup{\tup{q, f, g}, s, \gamma, \tup{q', f', g'}} \in \Delta^1_3$ if there exist $\tup{q, s, \alpha, q'} \in \Delta^1_2$ such that
		\begin{enumerate}
			\item $\tup{p', h} \in \mc{H}\paren{p, \alpha(x), f, g}$,
			\item $f'(p,x) = p'$,
			\item $g'(p,x,y) = \mc{S}^p_x(h)(y)$, and
			\item $\gamma\paren{Z^{p,x}_{y,k}} = \mc{A}^p_x(h)\paren{Z^{p,x}_{y,k}}$,
		\end{enumerate}
		\item $F^1_3(q, f, g) = \ms{Er}_Y \circ h \circ F^2_3(p)$, with $\tup{p, h} \in \mc{H}(p_0, F^1_2(q), f, g)$ chosen nondeterministically.
	\end{itemize}
\end{defi}

Besides accommodating nondeterminism, the use of $\mc{H}$ in \autoref{def:nsst_comp} does not change the essential nature of the composition construction.
In particular, the methods by which summaries and corresponding assignments are computed are identical in both \autoref{def:dsst_comp} and \autoref{def:nsst_comp}.
Only the way in which the computed summaries are grouped to form transitions in $\Delta^1_3$ differs.

\begin{rem}
	In order to minimize bookkeeping and notational complexity, the analyses in \autoref{subsec:analysis} and \autoref{subsec:copyless} are carried out for NSSTs using the maps $\mc{F}$ and $\mc{G}$ rather than $\mc{H}$, but with the assumption that the synchronization enforced by $\mc{H}$ holds implicitly.
\end{rem}

\subsection{Correctness and Size Bounds}
\label{subsec:analysis}
Suppose $\tup{s, t} \in \sem{T^1_2}$ and $\tup{t,u} \in \sem{T^2_3}$ such that there are corresponding runs $\rho = q_0 \xrightarrow[\alpha_1]{s_1} \dots \xrightarrow[\alpha_n]{s_n} q_n$ and $\sigma = p_0 \xrightarrow[\beta_1]{t_1}\dots \xrightarrow[\beta_m]{t_m} p_m$ in $T^1_2$ and $T^2_3$, respectively.
By \autoref{def:nsst_comp}, for every transition $\tup{q_{k-1}, s_k, \alpha_k, q_k} \in \Delta^1_2$ occurring in $\rho$, there exists a transition $\tup{\tup{q_{k-1}, f_{k-1}, g_{k-1}}, s_k, \gamma_k, \tup{q_k, f_k, g_k}} \in \Delta^1_3$.
Let $\tau$ be the corresponding run $\tup{q_0, f_0, g_0} \xrightarrow[\gamma_1]{s_1} \dots \xrightarrow[\gamma_n]{s_n} \tup{q_n, f_n, g_n}$ in $T^1_3$.
Also, let $\alpha = \comp^n_{k=1} \alpha_k$ and $\gamma = \comp^n_{k=1} \gamma_k$.

\begin{lem}
	\label{lemma:F_correctness}
	Suppose that $x \in X$ and $p \in P$.
	If $\sigma(p,x)$ is a run in $T^2_3$ on input $\mc{V}_\rho(x)$, starting from state $p$ and ending in state $p'$, then
	\begin{equation}
		\label{eq:F_correctness}
		f_n(p, x) = p'.
	\end{equation}
\end{lem}
\begin{proof}
	We proceed by induction on $n$, the length of $\rho$ and $\tau$.
	\begin{description}
		\item[Base case] \emph{\autoref{eq:F_correctness} holds when $n = 0$.}

			If $n = 0$, then $\rho = q_0$ and $\tau = \tup{q_0, f_0, g_0}$.
			Therefore, $\mc{V}_\rho(x) = \varepsilon$, for all $x \in X$, and so for all $p \in P$ and $x \in X$, we have $\sigma(p,x) = p = p'$.
			By \autoref{def:nsst_comp}, it holds that $f_0(p, x) = p$.

		\item[Inductive case] \emph{If \autoref{eq:F_correctness} holds for $n-1$, then it also holds for $n$.}

			By \autoref{def:nsst_comp}, $f_n(p, x) = \mc{F}(p, \alpha_n(x), f_{n-1})$, for all $\tup{p, x} \in P \times X$.
			Observe that $\mc{F}(p, \alpha_n(x), f_{n-1})$ partially simulates the transitions of $T^2_3$, by reading $\alpha_n(x)$ in sequence and keeping track of the current state.
			This simulation operates according to $\Delta^2_3$ when the next symbol of $\alpha_n(x)$ comes from $\Sigma_2$ and according to $f_{n-1}$ when the next symbol comes from $X$.
			In other words, for every state $p$, $\alpha_n(x)$ induces a sequence of states $p_1, p_2 \dots, p_{\card{\alpha_n(x)}}$, via $\mc{F}$, such that $p_1 = p$ and either
			\begin{enumerate}
				\item $\tup{p_k, s, \beta, p_{k+1}} \in \Delta^2_3$ for some $\beta \in B$, if the $k^{\tn{th}}$ symbol of $\alpha_n(x)$ is $s \in \Sigma_2$, or
				\item $f_{n-1}(p_k, x') = p_{k+1}$, if the $k^{\tn{th}}$ symbol of $\alpha_n(x)$ is $x' \in X$.
			\end{enumerate}
			The former case is correct by definition, while the latter follows from the inductive hypothesis.
			Consequently, we arrive at the equation
			\begin{equation*}
				f_n(p, x) = \mc{F}(p, \alpha_n(x), f_{n-1}) = p_{\card{\alpha_{n}(x)}} = p',
			\end{equation*}
			and we conclude that correctness of $f_{n-1}$ implies correctness of $f_n$. \qedhere
	\end{description}
\end{proof}

\begin{lem}
	\label{lemma:partial_commutativity}
	Suppose that $\gamma: \mc{Z} \to \brak{\mc{Z}, \Sigma_3}$, $h : Y \to \brak{Y \cup \mc{Z}, \Sigma_3}$, and $\phi(a) = a$ for any function $\phi$ and $a \notin \dom{\phi}$.
	For all $Z \in \mc{Z}$, the equation $\gamma(Z) = h \circ \gamma(Z) = \gamma \circ h(Z)$ holds.
\end{lem}
\begin{proof}
	Follows from the fact that $\dom{h} = Y$ is disjoint from $\im{\gamma} = \brak{\mc{Z}, \Sigma_3}$.
\end{proof}

\begin{lem}
	\label{lemma:G_correctness}
	Suppose that $x \in X$, $p \in P$, and $y \in Y$.
	If $\sigma(p,x)$ is a run in $T^2_3$ on input $\mc{V}_\rho(x)$, starting from state $p$, then
	\begin{equation}
		\label{eq:G_correctness}
		\mc{V}_{\sigma(p,x)}(y) = \ms{Er}_Z \circ \mc{V}_\tau \circ g_n(p, x, y).
	\end{equation}
\end{lem}
\begin{proof}
	We proceed by induction on $n$.
	\begin{description} 
		\item[Base case] \emph{\autoref{eq:G_correctness} holds when $n = 0$.}

			Because $\sigma(p, x) = p$, we have that $\mc{V}_{\sigma(p,x)}(y) = y$, for all $y \in Y$.
			By \autoref{def:nsst_comp} $g_0(p,x,y) = y$, and so $\mc{V}_\tau \circ g_0(p,x,y)= y$, since $\mc{V}_\tau$ is of type $\brak{\mc{Z}, \Sigma_3} \to \brak{\mc{Z}, \Sigma_3}$.
			Likewise, $\ms{Er}_Z$ does not erase variables from $Y$, so $\ms{Er}_Z \circ \mc{V}_\tau \circ g_0(p, x, y) = \mc{V}_{\sigma(p,x)}(y) = y$.

		\item[Inductive case] \emph{If \autoref{eq:G_correctness} holds for $n-1$, then it also holds for $n$.}

			Suppose that $\rho' = q_0 \xrightarrow[\alpha_1]{s_1} \dots \xrightarrow[\alpha_{n-1}]{s_{n-1}} q_{n-1}$, where $\rho = \rho' \xrightarrow[\alpha_n]{s_n} q_n$, and that $\tau' = \tup{q_0, f_0, g_0} \xrightarrow[\gamma_1]{s_1} \dots \xrightarrow[\gamma_{n-1}]{s_{n-1}} \tup{q_{n-1}, f_{n-1}, g_{n-1}}$ where $\tau = \tau' \xrightarrow[\gamma_n]{s_n} \tup{q_n, f_n, g_n}$.
			Further, suppose that $\sigma'(p,x)$ is a run in $T^2_3$ on $\mc{V}_{\rho'}(x)$, starting from state $p$.
			Let the last assignment of $\rho$ be $\alpha_n(x) = w_0 x_1 \dots x_\ell w_\ell$, and let $w^j_i$ denote the $j^\tn{th}$ symbol in $w_i$.
			The run $\sigma(p,x)$ in $T^2_3$ can be written as
			\begin{equation*}
				\sigma(p,x) = p \xrightarrow[\beta^1_0]{w^1_0} \dots p_{\card{w_0x_1 \dots w_{i-1}}} \xrightarrow[\mc{V}_{\sigma'(p_{|w_0x_1 \dots w_{i-1}|}, x_i)}]{x_i} p_{\card{w_0x_1 \dots w_{i-1} x_i}} \dots \xrightarrow[\beta^{\card{w_\ell}}_\ell]{w^{\card{w_\ell}}_\ell} p_{\card{\alpha_n(x)}}.
			\end{equation*}
			Consequently, we obtain the following derivation, which completes the proof.
			\begin{align}
				\label{eq:1}
				\mc{V}_{\sigma(p,x)} &= \comp^{\card{w_0}}_{k=1} \beta^k_0 \circ \comp^\ell_{i=1} \mc{V}_{\sigma'(p_{w_0x_1\dots w_{i-1}}, x_i)} \circ \comp^{\card{w_i}}_{j=1} \beta^i_j \\
				\label{eq:2}
				&= \comp^{\card{w_0}}_{k=1} \beta^k_0 \circ \comp^\ell_{i=1} \ms{Er}_Z \circ \mc{V}_{\tau'} \circ g_{n-1}(p_{w_0x_1\dots w_{i-1}}, x_i, \cdot) \circ \comp^{\card{w_i}}_{j=1} \beta^i_j \\
				\label{eq:3}
				&= \ms{Er}_Z \circ \mc{V}_{\tau'} \circ \comp^{\card{w_0}}_{k=1} \beta^k_0 \circ \comp^\ell_{i=1}  g_{n-1}(p_{w_0x_1\dots w_{i-1}}, x_i, \cdot) \circ \comp^{\card{w_i}}_{j=1} \beta^i_j \\
				\label{eq:4}
				&= \ms{Er}_Z \circ \mc{V}_{\tau'} \circ \mc{G}(p, \alpha_n(x), f_{n-1}, g_{n-1}) \\
				\label{eq:5}
				&= \ms{Er}_Z \circ \mc{V}_{\tau'} \circ \gamma_n \circ g_{n}(p, x, \cdot) \\
				\label{eq:6}
				&= \ms{Er}_Z \circ \mc{V}_\tau \circ g_{n}(p, x, \cdot)
			\end{align}
			From \autoref{eq:1}, we obtain \autoref{eq:2} from the inductive hypothesis.
			We get \autoref{eq:3} by applying \autoref{lemma:partial_commutativity}.
			From here, \autoref{eq:4} is obtained from the definition of $\mc{G}$ in \autoref{eq:ghat_dsst}.
			Next, \autoref{eq:5} follows from \autoref{eq:AS_combine} and the definitions of $\mc{S}$ and $\mc{A}$.
			Finally, \autoref{eq:6} is implied by the definition of valuation. \qedhere
	\end{description}
\end{proof}

\begin{thm}[Correctness]
	\label{theorem:correctness}
	If $T^1_3$ is the composition of $T^2_3$ and $T^1_2$, constructed according to \autoref{def:nsst_comp}, then $\sem{T^1_3} = \sem{T^2_3} \circ \sem{T^1_2}$.
\end{thm}
\begin{proof}
	Observe the following derivation:
	\begin{align}
		\label{eq:u1}
		\sem{T^1_3}(s) &= \ms{Er}_Z \circ \mc{V}_\tau \circ F^1_3(\tup{q_n, f_n, g_n}) \\
		\label{eq:u2}
		&= \ms{Er}_Z \circ \mc{V}_\tau \circ \ms{Er}_Y \circ \mc{G}(p_0, F^1_2(q_n), f_n, g_n) \circ F^2_3 \circ \mc{F}(p_0, F^1_2(q_n), f_n) \\
		\label{eq:u3}
		&= \ms{Er}_Y \circ \ms{Er}_Z \circ \mc{V}_\tau \circ \mc{G}(p_0, F^1_2(q_n), f_n, g_n) \circ F^2_3(p_m).
	\end{align}
	From \autoref{eq:u1}, we obtain \autoref{eq:u2} by applying \autoref{def:dsst_comp}.
	\autoref{eq:u3} follows from \autoref{eq:u2} via application of \autoref{lemma:F_correctness}.
	Considering the output of $T^1_2$ as the value of an extra variable $x_\tn{out}$, as in \cite{MuschollPuppis2019}.
	Letting $h = \mc{G}(p_0, F^1_2(q_n), f_n, g_n)$, we can rewrite \autoref{eq:u3} as
	\begin{equation}
		\label{eq:var}
		\sem{T^1_3}(s) = \ms{Er}_Y \circ \ms{Er}_Z \circ \mc{V}_\tau \circ \mc{A}^{p_0}_{x_{\tn{out}}}(h) \circ \mc{S}^{p_0}_{x_{\tn{out}}}(h) \circ F^2_3(p_m).
	\end{equation}
	Since $\sem{T^1_2}(s) = \ms{Er}_X \circ \mc{V}_\rho(x_\tn{out})$, by definition, applying \autoref{lemma:G_correctness} to \autoref{eq:var} yields
	\begin{equation*}
		\sem{T^1_3}(s) = \ms{Er}_Y \circ \mc{V}_{\sigma} \circ F^2_3(p_m) = \sem{T^2_3}(t). \qedhere
	\end{equation*}
\end{proof}

\begin{thm}
	Suppose that $T^1_3$ is the composition of $T^2_3$ and $T^1_2$ constructed according to \autoref{def:nsst_comp}.
	If $T^1_2$ has $k$ states and $\ell$ variables and $T^2_3$ has $n$ states and $m$ variables, then $T^1_3$ has a state space of size $O(k n^{\ell n} ((2\ell nm + m)!)^{\ell nm})$ and $2\ell nm$ variables.
\end{thm}
\begin{proof}
	By \autoref{def:dsst_comp}, the state space of $R$ of $T^1_3$ is comprised by the product $Q \times \bb{F} \times \bb{G}$ where $\bb{F} = (P \times X) \to P$ and $\bb{G} = (P \times X \times Y) \to \brak{Y \cup \mc{Z}}$.
	Therefore, we have
	\begin{gather*}
		\card{\bb{F}} = \card{P}^{\card{P \times X}} = n^{\ell n} \qquad\tn{ and }\qquad \card{\bb{G}} = \card{\brak{Y \cup \mc{Z}}}^{\card{P \times X \times Y}} = \card{\brak{Y \cup \mc{Z}}}^{\ell nm}.
	\end{gather*}
	It is clear from \autoref{def:dsst_comp} that $\card{\mc{Z}} = 2\ell nm$.
	Applying \autoref{proposition:copyless_card_bound} yields
	\begin{equation*}
		\card{\brak{Y \cup \mc{Z}}} = \floor*{e (m + 2\ell nm)!}.
	\end{equation*}
	As a result, we get that
	\begin{equation*}
		\card{R} = k n^{\ell n} \floor*{e (m + 2\ell nm)!}^{\ell nm} = O\paren{k n^{\ell n} ((2\ell nm + m)!)^{\ell nm}}. \qedhere
	\end{equation*}
\end{proof}

\subsection{Establishing the Diamond-Free Property}
\label{subsec:copyless}
We define two binary relations over the $\mc{Z}$ variables of $T^1_3$:
\begin{equation*}
	Z^{p,x}_{y,b} \eqx Z^{x', p'}_{y', b'} \iff x = x' \qquad\tn{ and }\qquad	Z^{p,x}_{y,b} \equiv Z^{x', p'}_{y', b'} \iff x = x' \land p = p'.
\end{equation*}
It is easy to see that $\eqx$ and $\equiv$ are equivalence relations.
The relation $\eqx$ partitions $\mc{Z}$ into $|X|$ classes, each of size $2|Y {\times} P|$.
Similarly, the relation $\equiv$ partitions $\mc{Z}$ into  $|X \times P|$ classes each of size $2|Y|$.
We write $Z|_{\eqx} = x$ to denote that $Z$ belongs to the class indexed by $x$ and $Z|_\equiv = \tup{p, x}$ to denote that the equivalence class of $Z$ is indexed by $\tup{p, x}$.
We next utilize these relations and the structure they impose on $\mc{Z}$ to show that the composite transducer is bounded-copy.
In the following, $\alpha$ is always an assignment in $T^1_3$, and $Z, Z', Z_1,$ etc. will refer to generic variables from $\mc{Z}$.

\begin{lem}
	\label{lemma:1g}
	For any state $\tup{q, f, g} \in R$, variable $Z \in \mc{Z}$, and $\tup{p, x, y} \in P \times X \times Y$, it holds that $\card{g(p, x, y)}_Z \leq 1$ and if $Z \in g(p, x, y)$, then $Z|_\equiv = \tup{p, x}$.
\end{lem}
\begin{proof}
	By definition of the initial shape summary $g_0$ it is clear that $\card{g_0(p, x, y)}_Z = 0$, regardless of $p,x,y,Z$.
	Any other shape summary occurring in $T^1_3$ is constructed by the map $\mc{S}$ which is defined, in \autoref{eq:shape}, such that no variable from $\mc{Z}$ occurs twice in the image of a single element of the domain.
	Additionally, it is easy to see from the same definition in \autoref{eq:shape} that any $Z \in \mc{S}(p, x, h)(y)$ must be indexed by the pair $\tup{p, x}$, invariant of $h$ and $y$, thus implying that $Z|_\equiv = \tup{p, x}$.
\end{proof}

\begin{lem}
	\label{lemma:zneqx}
	Let $\alpha : X \to \brak{X, \Sigma_2}$ be arbitrary and suppose that $Z_1 \in \mc{G}(p_1, \alpha(x_1), f, g)(y_1)$ and $Z_2 \in \mc{G}(p_2, \alpha(x_2), f, g)(y_2)$.
	If $x_1 \neq x_2$, then $Z_1 \not\eqx Z_2$.
\end{lem}
\begin{proof}
	If $Z_1 \eqx Z_2$, then there must be a unique variable $x$ such that $\mc{G}(p_1, \alpha(x_1), f, g)$ uses a shape summary $g^{p_1'}_x$ and $\mc{G}(p_2, \alpha(x_2), f, g)$ uses a shape summary $g^{p_2'}_x$ for some states $p'_1, p'_2$.
	This implies that $x \in \alpha(x_1)$ and $x \in \alpha(x_2)$ which contradicts our assumption that $\alpha$ is copyless.
	Consequently, we conclude that the pair of inclusions $Z_1 \in \mc{G}(p_1, \alpha(x_1), f, g)(y_1)$ and $Z_2 \in \mc{G}(p_2, \alpha(x_2), f, g)(y_2)$ imply that $Z_1 \not\eqx Z_2$.
\end{proof}

\begin{lem}
	\label{lemma:1G}
	For any state $\tup{q, f, g} \in R$, copyless assignment $\alpha : X \to \brak{X, \Sigma_2}$, variable $Z \in \mc{Z}$, and $\tup{p, x, y} \in P \times X \times Y$ it holds that $\card{\mc{G}(p, \alpha(x), f, g)(y)}_Z \leq 1$.
\end{lem}
\begin{proof}
	We proceed by induction on $\alpha(x)$.
\begin{description}
    \item[Base case] $\card{\mc{G}(p, \varepsilon, f, g)}_Z \leq 1$.

	By definition of $\mc{G}$, in \autoref{eq:ghat_dsst}, it follows that $\mc{G}(p, \varepsilon, f, g)(y) = y$ when $\alpha(x) = \varepsilon$.
	Since no variables from $\mc{Z}$ ever occur in such a string, $\card{\mc{G}(p, \varepsilon, f, g)(y)}_Z = 0 \leq 1$.

	\item[Inductive case] \emph{If $\card{\mc{G}(p, \mb{x}, f, g)(y)}_Z \leq 1$ and either $\alpha(x) = \mb{x} s$ or $\alpha(x) = \mb{x} x'$, then} 
	\begin{equation*}
		\card{\mc{G}(p, \alpha(x), f, g)(y)}_Z \leq 1.
	\end{equation*}

	The disjunction contained in the inductive hypothesis leads us to consider two cases.

	\begin{enumerate}
		\item 
		Suppose that $\alpha(x) = \mb{x} s$.
		By applying the definition of $\mc{G}$ we obtain the equation
		\begin{equation*}
			\mc{G}(p, \alpha(x), f, g) = \mc{G}(p, \mb{x}, f, g) \circ \beta,
		\end{equation*}
		where $(\mc{F}(p, \mb{x}, f), s, \beta, p') \in \Delta^2_3$.
		Since the type of $\beta$ is $Y \to \brak{Y, \Sigma_3}$, composing this assignment with $\mc{G}(p, \mb{x}, f, g)$ does not introduce or remove any $\mc{Z}$ variables from the resulting strings, we get
		\begin{equation}
			\label{eq:a}
			\card{\mc{G}(p, \mb{x} s, f, g)(y)}_Z =  \card{\mc{G}(p, \mb{x}, f, g) \circ \beta(y)}_Z = \card{\mc{G}(p, \mb{x}, f, g)(y)}_Z.
		\end{equation}
		Assuming the inductive hypothesis, we have $\card{\mc{G}(p, \mb{x}, f, g)(y)}_Z \leq 1$, which, in combination with \autoref{eq:a}, implies the inequality $\card{\mc{G}(p, \mb{x} s, f, g)(y)}_Z \leq 1$.
		\item 
		Otherwise, suppose that $\alpha(x) = \mb{x} x'$.
		Then, applying the definition of $\mc{G}$ implies that
		\begin{equation*}
			\mc{G}(p, \beta(x), f, g)(y) = \mc{G}(p, \mb{x}, f, g) \circ g(\mc{F}(p, \mb{x}, f), x', y)
		\end{equation*}
		holds for every $y \in Y$.
		By assumption, we know that $\card{\mc{G}(p, \mb{x}, f, g)(y)}_Z \leq 1$, and, applying \autoref{lemma:1g}, we obtain the inequality $\card{g(\mc{F}(p, \mb{x}, f), x', y)}_Z \leq 1$.
		However, $Z \in \mc{G}(p, \mb{x}, f, g)(y)$ and $Z \in g(\mc{F}(p, \mb{x}, f), x', y)$ cannot hold simultaneously, since this would imply that $\card{\alpha(x)}_{x'} = 2$ thereby contradicting the assumption that $\alpha$ is copyless.
		This mutual exclusivity may be restated quantitatively as
		\begin{equation*}
			\card{\mc{G}(p, \mb{x}, f, g)(y)}_Z + \card{g(\mc{F}(p, \mb{x}, f), x', y)}_Z \leq 1,
		\end{equation*}
		and this implies $\card{\mc{G}(p, \mb{x} x', f, g)(y)}_Z \leq 1$.\qedhere
	\end{enumerate}
	\end{description}
\end{proof}

\begin{lem}
	\label{lemma:GSumY}
	For any state $\tup{q, f, g} \in R$, copyless assignment $\alpha : X \to \brak{X, \Sigma_2}$, variable $Z \in \mc{Z}$, and $\tup{p, x} \in P \times X$, it holds that $\sum_Y \card{\mc{G}(p, \alpha(x), f, g)(y)}_Z \leq 1$.
\end{lem}
\begin{proof}
	Suppose that $\tup{q, s, \alpha, q'} \in \Delta^1_2$ is the corresponding transition in $T^1_2$.
	If both the inclusions $Z \in \mc{G}(p, \alpha(x), f, g)(y_1)$ and $Z \in \mc{G}(p, \alpha(x), f, g)(y_2)$ hold, then there must exist $\mb{x}, \mb{x}' \in (X \cup \Sigma_2)^*$ and $x' \in X$ such that $\mb{x} x' \mb{x}' = \alpha(x)$ and $Z \in g(\mc{F}(p, \mb{x}, f), x', y_3)$, for some variable $y_3$, occurring both $\mc{G}(p, \mb{x}, f, g)(y_1)$ and $\mc{G}(p, \mb{x}, f, g)(y_2)$.
	However, we know that compositions of copyless assignments, such as $\mc{G}(p, \mb{x}, f, g)$, remain copyless.
	Thus, if $y_3 \in \mc{G}(p, \mb{x}, f, g)(y_1)$ and $y_3 \in \mc{G}(p, \mb{x}, f, g)(y_2)$ both hold, there must exist a copyful assignment in $T^2_3$.
	This contradicts our initial assumption that $T^2_3$ is copyless, so we infer that a single variable $Z$ cannot occur in more than one element of the set $\set{\mc{G}(p, \alpha(x), f, g)(y) : y \in Y}$.
	This fact, in combination with \autoref{lemma:1G}, which asserts that $Z$ can occur at most once in any element of this set, allows us to conclude that $\sum_{y \in Y} \card{\mc{G}(p, \alpha(x), f, g)(y)}_Z \leq 1$.
\end{proof}

\begin{lem}
	\label{lemma:XPY}
	Consider $Z \in \mc{Z}$ and $\alpha : X \to \brak{X, \Sigma_2}$.
	If $Z \in \mc{G}(p_1, \alpha(x_1), f, g)(y_1)$ and $Z \in \mc{G}(p_2, \alpha(x_2), f, g)(y_2)$, then $x_1 = x_2$ and either (i) $p_1 = p_2$ and $y_1 = y_2$ or (ii) $p_1 \neq p_2$.
\end{lem}
\begin{proof}
	Since $Z \eqx Z$, \autoref{lemma:zneqx} implies that $x_1 = x_2$, we are left with two cases.
	\begin{enumerate}
		\item
		First, suppose that $p_1 = p_2$.
		If $y_1 \neq y_2$, then we have that $Z \in \mc{G}(p, \alpha(x), f, g)(y_1)$ and $Z \in \mc{G}(p, \alpha(x), f, g)(y_2)$.
		It implies that 
		\begin{equation*}
			\card{\mc{G}(p, \alpha(x), f, g)(y_1)}_Z + \card{\mc{G}(p, \alpha(x), f, g)(y_2)}_Z = 2,
		\end{equation*}
		and thus contradicts \autoref{lemma:GSumY}.
		Therefore, it follows that $p_1 = p_2$ implies $y_1 = y_2$.
		\item
		Otherwise, suppose that $p_1 \neq p_2$.
		If $y_1 = y_2$, it follows that $Z \in \mc{G}(p_1, \alpha(x), f, g)(y)$ and $Z \in \mc{G}(p_2, \alpha(x), f, g)(y)$.
		Conversely, if $y_1 \neq y_2$, then $Z \in \mc{G}(p_1, \alpha(x), f, g)(y_2)$ and $Z \in \mc{G}(p_2, \alpha(x), f, g)(y_1)$.
		Neither possibility raises a contradiction, so the inequality $p_1 \neq p_2$ is independent of the relationship between $y_1$ and $y_2$. \qedhere
	\end{enumerate}
\end{proof}

\begin{lem}
	\label{lemma:out}
	Suppose that $\gamma : \mc{Z} \to \brak{\mc{Z}, \Sigma_3}$ and $Z,Z_1,Z_2 \in \mc{Z}$.
	If $Z \in \gamma(Z_1)$ and $Z \in \gamma(Z_2)$, then $Z_1 \eqx Z_2$ and either $Z_1 = Z_2$ or $Z_1 \not\equiv Z_2$.
\end{lem}
\begin{proof}
	If $Z \in \gamma(Z_1)$ and $Z \in \gamma(Z_2)$, then there exist pairs $\tup{p_1, x_1}$ and $\tup{p_2, x_2}$ such that we have the inclusions $Z \in \mc{A}^{p_1}_{x_1}(\mc{G}(p_1, \alpha(x_1), f, g))(Z_1)$ and $Z \in \mc{A}^{p_2}_{x_2}(\mc{G}(p_2, \alpha(x_2), f, g))(Z_2)$.
	From \autoref{eq:assign}, it follows that if these memberships hold, then the pair of memberships $Z \in \mc{G}(p_1, \alpha(x_1), f, g)(y_1)$ and $Z \in \mc{G}(p_2, \alpha(x_2), f, g)(y_2)$ also hold, for some $y_1, y_2 \in Y$.
	Now, \autoref{lemma:XPY} entails that either $x_1 = x_2$, $p_1 = p_2$, $y_1 = y_2$, or $x_1 = x_2$ and $p_1 \neq p_2$.
	In either case, $Z_1 \eqx Z_2$.
	For the former case, we have equality of the states and $Y$ variables, which, in combination with \autoref{lemma:GSumY}, implies $Z_1 = Z_2$.
	In the latter case, $Z_1$ and $Z_2$ are not indexed by the same state, so $Z_1 \not\equiv Z_2$.
\end{proof}

\begin{lem}
	\label{lemma:in}
	For $\gamma : \mc{Z} \to \brak{\mc{Z}, \Sigma_3}$, if $Z_1 \in \gamma(Z)$ and $Z_2 \in \gamma(Z)$, then $Z_1 \equiv Z_2$ or $Z_1 \not\eqx Z_2$.
\end{lem}
\begin{proof}
	If $Z_1, Z_2 \in \gamma(Z)$, then $Z_1, Z_2 \in \mc{G}(p, \alpha(x), f, g)(y)$ for some $y$.
	Both variables must be introduced by a shape summary constructed by $\mc{G}(p, \alpha(x), f, g)$.
	This yields two cases: (1) both variables belong to a summary mapping $g^{p'}_{x'}$ with common parameters from $P$ and $X$, or (2) each variable belongs to summary mappings $g^{p_1}_{x_1}$ and $g^{p_2}_{x_2}$ with distinct parameters.
	When $Z_1, Z_2$ both occur in some element of $\bigcup_{y \in Y} g(p', x', y)$, then $Z_1 \equiv Z_2$ by \autoref{lemma:1g}.
	Otherwise, when $Z_1$ occurs in an element of $\bigcup_{y \in Y} g(p_1, x_1, y)$ and $Z_2$ occurs in some element of $\bigcup_{y \in Y} g(p_2, x_2, y)$, it must hold that $x_1 \neq x_2$ because $\alpha$ is copyless.
	Since the $\mc{Z}$ variables in question cannot be indexed by the same $X$ variable, we conclude that $Z_1 \not\eqx Z_2$. 
\end{proof}

\begin{thm}
	\label{theorem:diamond-free}
	If $T^1_3$ is the composition of $T^2_3$ and $T^1_2$ constructed according to \autoref{def:nsst_comp}, then $T^1_3$ is diamond-free.
\end{thm}
\begin{proof}
	\autoref{lemma:in} and \autoref{lemma:out} imply that any pair of copies of a single variable never occur in a common variable later in the computation.
	Therefore, it is impossible for a diamond to occur in any flow graph of a run in $T^1_3$.
\end{proof}

\section{Removing Copies from Diamond-Free NSSTs}
\label{sec:toFNSST}
In this section, provide a procedure that produces an equivalent copyless NSST when given a diamond-free NSST.
The SST $T_3$ from \autoref{fig:Alur_comp}, also displayed in \autoref{fig:diaomondless_DSST} along with the flow graphs of its assignments, will be used as a running example to illustrate the argument.

\begin{figure}[b]
	\begin{subfigure}{0.4\linewidth}
		\centering
		\begin{tikzpicture}
			\node[state,initial] (0) {$r_0$};
			\node[state,accepting] (1)[right = 1.5cm of 0] {$r_1$};
			\node (F)[right = 1cm of 1] {};

			\path[->] (0) edge node[above] {$a \backslash \gamma_1$} (1);
			\path[->] (1) edge[loop above] node {$a \backslash \gamma_2$}  (1);
			\path[->] (1) edge node[above] {$z_1z_2$} (F);
		\end{tikzpicture}
		\subcaption{$T_3$}
		\label{subfig:simple_T3}
	\end{subfigure}%
	\begin{subfigure}{0.6\linewidth}
		\centering
		\begin{tikzpicture}
			\node (A) [draw  = gray, label = $G(\gamma_1)$, inner sep = 2pt, rounded corners] {
				\begin{tikzpicture}[node distance = 1cm]
					\node (a1) {$z_1$};
					\node (a2) [right of = a1] {$z_2$};
					\node (a3) [right of = a2] {$z_3$};
					\node (a4) [right of = a3] {$z_4$};
					\node (1a) [below = 0.5cm of a1] {$z_1$};
					\node (2a) [right of = 1a] {$z_2$};
					\node (3a) [right of = 2a] {$z_3$};
					\node (4a) [right of = 3a] {$z_4$};
				\end{tikzpicture}
			};
			\node (B) [right = 0.5cm of A, draw  = gray, label = $G(\gamma_2)$, inner sep = 2pt, rounded corners] {
				\begin{tikzpicture}[node distance = 1cm]
					\node (b1) {$z_1$};
					\node (b2) [right of = b1] {$z_2$};
					\node (b3) [right of = b2] {$z_3$};
					\node (b4) [right of = b3] {$z_4$};
					\node (1b) [below = 0.5cm of b1] {$z_1$};
					\node (2b) [right of = 1b] {$z_2$};
					\node (3b) [right of = 2b] {$z_3$};
					\node (4b) [right of = 3b] {$z_4$};

					\path[->] (b3) edge (1b);
					\path[->] (b3) edge (3b);
					\path[->] (b4) edge (2b);
					\path[->] (b4) edge (4b);
				\end{tikzpicture}
			};
		\end{tikzpicture}
		\subcaption{Flow graphs of $\gamma_1$ and $\gamma_2$.}
		\label{subfig:T3_flows}
	\end{subfigure}
	\caption{The SST $T_3$ and flow graphs for each of its assignments.}
	\label{fig:diaomondless_DSST}
\end{figure}

\begin{figure}
	\centering
	\resizebox{\linewidth}{!}{
		\begin{tikzpicture}[> = Stealth]
			\node (B) [draw  = gray, rounded corners, inner sep=2pt, label={180:$G(\gamma_2)$}] {
				\begin{tikzpicture}[node distance = 1cm]
					\node (b1) {$z_1$};
					\node (b2) [right of = b1] {$z_2$};
					\node (b3) [right of = b2] {$z_3$};
					\node (b4) [right of = b3] {$z_4$};
					\node (1b) [below = 0.5cm of b1] {$z_1$};
					\node (2b) [right of = 1b] {$z_2$};
					\node (3b) [right of = 2b] {$z_3$};
					\node (4b) [right of = 3b] {$z_4$};

					\path[->] (b3) edge (1b);
					\path[->] (b3) edge (3b);
					\path[->] (b4) edge (2b);
					\path[->] (b4) edge (4b);
				\end{tikzpicture}
			};
			\node (C) [below left = 1.25cm and 0.25cm of B, draw  = gray, xshift=1cm, inner sep=2pt, rounded corners] {
				\begin{tikzpicture}[node distance = 1cm]
					\node (b1) {$z_1$};
					\node (b2) [right of = b1] {$z_2$};
					\node (b3) [right of = b2] {$z_3$};
					\node (b4) [right of = b3] {$z_4$};
					\node (1b) [below = 0.5cm of b1] {$z_1$};
					\node (2b) [right of = 1b] {$z_2$};
					\node (3b) [right of = 2b] {$z_3$};
					\node (4b) [right of = 3b] {$z_4$};

					\path[->] (b3) edge (1b);
					\path[->] (b4) edge (2b);
					\path[->] (b4) edge (4b);
				\end{tikzpicture}
			};
			\node (D) [below right = 1.25cm and 0.25cm of B, draw  = gray, xshift=-1cm, inner sep=2pt, rounded corners] {
				\begin{tikzpicture}[node distance = 1cm]
					\node (b1) {$z_1$};
					\node (b2) [right of = b1] {$z_2$};
					\node (b3) [right of = b2] {$z_3$};
					\node (b4) [right of = b3] {$z_4$};
					\node (1b) [below = 0.5cm of b1] {$z_1$};
					\node (2b) [right of = 1b] {$z_2$};
					\node (3b) [right of = 2b] {$z_3$};
					\node (4b) [right of = 3b] {$z_4$};

					\path[->] (b3) edge (3b);
					\path[->] (b4) edge (2b);
					\path[->] (b4) edge (4b);
				\end{tikzpicture}
			};
			\node (B1) [below left = 1.25cm and 0.25cm of C, draw  = gray, label={270:$G(\gamma_{2,1})$}, xshift=0.5cm, inner sep=2pt, rounded corners] {
				\begin{tikzpicture}[node distance = 1cm]
					\node (b1) {$z_1$};
					\node (b2) [right of = b1] {$z_2$};
					\node (b3) [right of = b2] {$z_3$};
					\node (b4) [right of = b3] {$z_4$};
					\node (1b) [below = 0.5cm of b1] {$z_1$};
					\node (2b) [right of = 1b] {$z_2$};
					\node (3b) [right of = 2b] {$z_3$};
					\node (4b) [right of = 3b] {$z_4$};

					\path[->] (b3) edge (1b);
					\path[->] (b4) edge (2b);
				\end{tikzpicture}
			};
			\node (B2) [below right = 1.25cm and 0.25cm of C, draw  = gray, label={270:$G(\gamma_{2,2})$}, xshift=-3.25cm, inner sep=2pt, rounded corners] {
				\begin{tikzpicture}[node distance = 1cm]
					\node (b1) {$z_1$};
					\node (b2) [right of = b1] {$z_2$};
					\node (b3) [right of = b2] {$z_3$};
					\node (b4) [right of = b3] {$z_4$};
					\node (1b) [below = 0.5cm of b1] {$z_1$};
					\node (2b) [right of = 1b] {$z_2$};
					\node (3b) [right of = 2b] {$z_3$};
					\node (4b) [right of = 3b] {$z_4$};

					\path[->] (b3) edge (1b);
					\path[->] (b4) edge (4b);
				\end{tikzpicture}
			};
			\node (B3) [below left = 1.25cm and 0.25cm of D, draw  = gray, label={270:$G(\gamma_{2,3})$}, xshift=3.25cm, inner sep=2pt, rounded corners] {
				\begin{tikzpicture}[node distance = 1cm]
					\node (b1) {$z_1$};
					\node (b2) [right of = b1] {$z_2$};
					\node (b3) [right of = b2] {$z_3$};
					\node (b4) [right of = b3] {$z_4$};
					\node (1b) [below = 0.5cm of b1] {$z_1$};
					\node (2b) [right of = 1b] {$z_2$};
					\node (3b) [right of = 2b] {$z_3$};
					\node (4b) [right of = 3b] {$z_4$};

					\path[->] (b3) edge (3b);
					\path[->] (b4) edge (2b);
				\end{tikzpicture}
			};
			\node (B4) [below right = 1.25cm and 0.25cm of D, draw=gray, label={270:$G(\gamma_{2,4})$}, xshift=-0.5cm, inner sep=2pt, rounded corners] {
				\begin{tikzpicture}[node distance = 1cm]
					\node (b1) {$z_1$};
					\node (b2) [right of = b1] {$z_2$};
					\node (b3) [right of = b2] {$z_3$};
					\node (b4) [right of = b3] {$z_4$};
					\node (1b) [below = 0.5cm of b1] {$z_1$};
					\node (2b) [right of = 1b] {$z_2$};
					\node (3b) [right of = 2b] {$z_3$};
					\node (4b) [right of = 3b] {$z_4$};

					\path[->] (b3) edge (3b);
					\path[->] (b4) edge (4b);
				\end{tikzpicture}
			};

			\path[->] (B) edge (C);
			\path[->] (B) edge (D);
			\path[->] (C) edge (B1);
			\path[->] (C) edge (B2);
			\path[->] (D) edge (B3);
			\path[->] (D) edge (B4);
		\end{tikzpicture}
	}
	\caption{\decomp{} applied to the copyful assignment $\gamma_2$ of $T_3$.}
	\label{fig:illdecompose}
\end{figure}
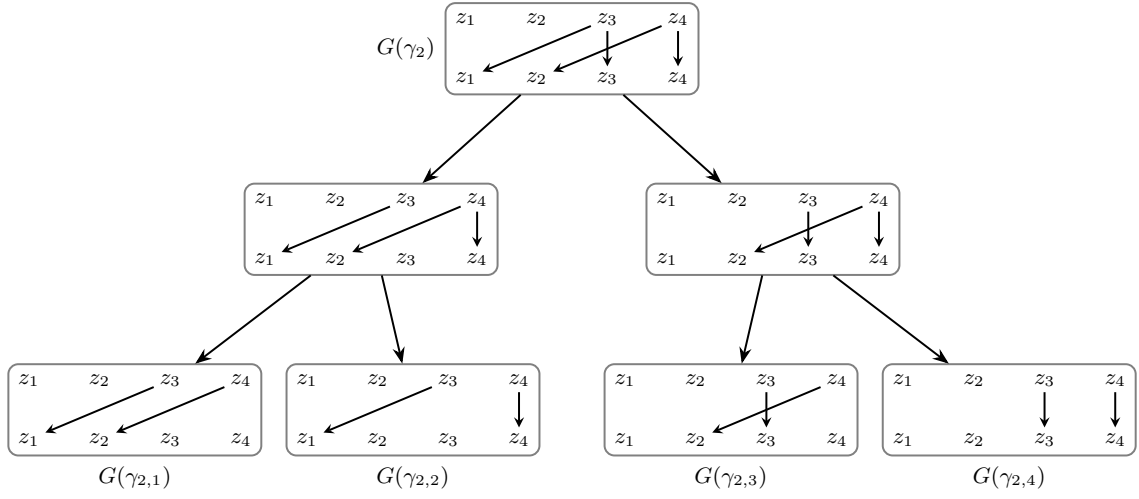

\paragraph{Assignment Decomposition.}
The first portion of the conversion procedure involves decomposing each copyful assignment $\alpha$ in the given diamond-free NSST into a collection of copyless assignments that ``cover'' $\alpha$ in the sense that the union of these copyless mappings coincides exactly with the mapping $\alpha$.
This decomposition subroutine, \decomp{}, is given in \autoref{alg:decompose}.
\autoref{fig:illdecompose} provides an illustration of the execution of \decomp{} when given the copyful assignment $\gamma_2$ of $T_3$ as input.

\begin{algorithm}
	\caption{Decomposition of a copyful assignment into copyless assignments.}
	\label{alg:decompose}
	\DontPrintSemicolon
	\SetKwFunction{D}{Decompose}
	\SetKwProg{Fn}{function}{}{}
	\Fn{\D{$\alpha$}}{
		\If{$\alpha$ is copyless \label{cpless_check1}}{
			\KwRet{$\set{\alpha}$}
		}
		$D \gets \emptyset$\;
		Choose $x \in X$ with distinct $y,z \in X$ such that $x \in \alpha(y)$ and $x \in \alpha(z)$\;
		\ForEach{$y \in X$ such that $x \in \alpha(y)$}{
			\ForEach{$z \in X \smallsetminus \set{y}$ such that $x \in \alpha(z)$}{
				Let $\beta$ be a new assigment that is identical to $\alpha$\;
				Set $\beta(z) = \varepsilon$\;
				\eIf{$\beta$ is copyless \label{cpless_check2}}{
					$D \gets D \cup \set{\beta}$%
				}{
					$D \gets D$ $\cup$ \D{$\beta$}%
				}
			}
		}
		\KwRet{$D$}\;
	}
\end{algorithm}

\begin{lem}
	\label{lemma:decomp}
	If $\alpha$ is diamond-free, then each $\alpha' \in \decomp{}(\alpha)$ is copyless and
	\begin{equation*}
		\alpha = \bigcup\limits_{\alpha' \in \decomp{}(\alpha)} \alpha'.
	\end{equation*}
\end{lem}
\begin{proof}
	Each $\alpha_k$ being copyless follows from the fact that the conditionals at lines \ref{cpless_check1} and \ref{cpless_check2} of \autoref{alg:decompose} ensure only copyless assignments are added to $D$.
	If the conditional at line \ref{cpless_check1} is entered then $\set{\alpha}$ is returned and $\alpha = \alpha$, thus satisfying the second assertion of the lemma.
	Otherwise, a variable $x$ is selected such that $x$ is copied by $\alpha$.
	The subsequent loop iterates through the possible target variables of for $x$, fixes one such $y$ at each iteration, creates a new assignment $\beta$ as an identical copy of $\alpha$, and then deletes $\beta(z)$ for all other variables $z$ such that $x \in \alpha(z)$.
	Thus, every mapping in the assignment $\alpha$ is present in $\beta$ for at least one iteration of the loop.
	Furthermore, this process either adds each $\beta$ into $D$ or recurses with $\beta$ as a parameter.
	The former guarantees that at least one assignment in $D$ includes any given mapping from $\alpha$.
	In the latter case, a mapping will not be removed later on, since the variable $x$ will no longer be copied by any assignment $\beta$ passed recursively to \decomp{}.
	Therefore, every piece of $\alpha$ occurs in some assignment in $D$ and $\alpha = \bigcup_{\alpha' \in \decomp{}(\alpha)} \alpha'$.
\end{proof}


\begin{defi}[Diamond-Free $\to$ Copyless]
	\label{def:diamond-free_to_copyless}
	Suppose that $T = \tup{\Sigma, \Lambda, Q, q_0, X, A, \Delta, F}$ is a diamond-free NSST.
	Let $T' = \tup{\Sigma, \Lambda, Q', q'_0, X, A', \Delta', F'}$ be a copyless NSST where
	\begin{itemize}
		\item $Q' = Q \times 2^X$ and $q'_0 = \tup{q_0, \emptyset}$;
		\item $A' = \bigcup\limits_{\alpha \in A} \decomp{}(\alpha)$;
		\item $\tup{\tup{p, Y}, s, \beta, \tup{q, Z}} \in \Delta'$ if the following conditions both hold:
		\begin{enumerate}
			\item no pair $x,y \in X$ exists such that $y \in Y$ and $y \in \beta(x)$,
			\item $\tup{p, s, \alpha, q} \in \Delta$ such that $\beta \in \decomp{}(\alpha)$ and
			\begin{equation*}
				y \in Z \iff \exists x \in X.\: x \in \alpha(y) \land x \notin \beta(y);
			\end{equation*}
		\end{enumerate}
		\item $F'(\tup{q, Y}) = F(q)$ if $x \notin F(q)$, for all $x \in Y$.
	\end{itemize}
\end{defi}

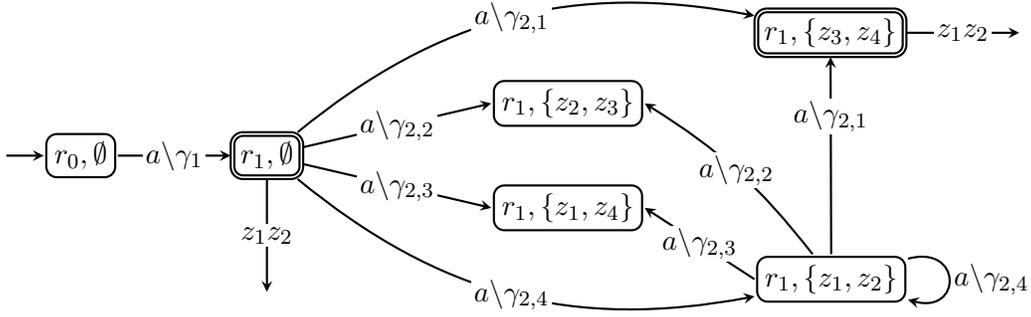
\begin{figure}
	\centering
	\begin{tikzpicture}
		\tikzset{
			state/.style = {rectangle, rounded corners, draw = black, minimum size = 1em, text centered, inner sep = 3pt},
			every node/.style = {inner sep = 1pt, outer sep = 1pt, minimum size = 0pt, fill = white},
			every loop/.style = {max distance = 0.75cm}
		}
		\node[state,initial] (0) {$r_0, \emptyset$};
		\node[state,accepting] (00)[right = 1.5cm of 0] {$r_1, \emptyset$};
		\node[state,accepting] (1)[above right = 1cm and 6cm of 00] {$r_1, \set{z_3, z_4}$};
		\node[state] (2)[above right = 0.05cm and 2.5cm of 00] {$r_1, \set{z_2, z_3}$};
		\node[state] (3)[below right = 0.05cm and 2.5cm of 00] {$r_1, \set{z_1, z_4}$};
		\node[state] (4)[below right = 1cm and 6cm of 00] {$r_1, \set{z_1, z_2}$};
		\node (F)[right = 1.5cm of 1] {};
		\node (F0)[below = 1.5cm of 00] {};

		\path[->] (0) edge node {$a \backslash \gamma_1$} (00);
		\path[->] (00) edge[bend left = 25] node {$a \backslash \gamma_{2,1}$} (1);
		\path[->] (00) edge node {$a \backslash \gamma_{2,2}$} (2.west);
		\path[->] (00) edge node {$a \backslash \gamma_{2,3}$} (3.west);
		\path[->] (00) edge[bend right = 25] node {$a \backslash \gamma_{2,4}$} (4);

		\path[->] (4) edge[loop right] node {$a \backslash \gamma_{2,4}$} (4);
		\path[->] (4) edge node[pos = 0.7] {$a \backslash \gamma_{2,1}$} (1);
		\path[->] (4) edge[bend right = 10] node {$a \backslash \gamma_{2,2}$} (2.east);
		\path[->] (4.west) edge node {$a \backslash \gamma_{2,3}$} (3.east);
		
		\path[->] (1) edge node {$z_1z_2$} (F);
		\path[->] (00) edge node {$z_1z_2$} (F0);
	\end{tikzpicture}
	\caption{A copyless functional NSST equivalent to $T_3$ from \autoref{fig:Alur_comp} and constructed according to \autoref{def:diamond-free_to_copyless}.
		Note that we omit disconnected states from the diagram.}
\end{figure}

In the sequel, suppose that $T = \tup{\Sigma, \Lambda, Q, q_0, X, A, \Delta, F}$ is a diamond-free NSST.
Additionally suppose that $T' = \tup{\Sigma, \Lambda, Q', q'_0, X, A', \Delta', F'}$ is the copyless NSST constructed according to \autoref{def:diamond-free_to_copyless}.

\begin{lem}
	\label{lemma:big_decomp}
	If $\rho$ is an arbitrary run $q_0 \xrightarrow[\alpha_1]{w_1} \dots \xrightarrow[\alpha_n]{w_n} q_n$ in $T$ and $\mc{R}(\rho)$ is the set of all runs of the form $\tup{q_0, \emptyset} \xrightarrow[\beta_1]{w_1} \dots \xrightarrow[\beta_n]{w_n} \tup{q_n, Y_n}$ in $T'$, then
	\begin{equation*}
		\mc{V}_\rho = \bigcup_{\rho' \in \mc{R}(\rho)} \mc{V}_{\rho'}
	\end{equation*}
\end{lem}
\begin{proof}
	We proceed by induction on the run $\rho$.
	\begin{description}
		\item[Base case] \emph{$\rho$ is the empty run.}
		
			In this case, $\mc{R}(\rho)$ is a singleton set containing the empty run and thus $\mc{V}_\rho = \bigcup_{\rho' \in \mc{R}(\rho)} \mc{V}_{\rho'}$.

		\item[Inductive case] \emph{$\sigma = \rho \xrightarrow[\alpha]{s} r$ is a run in $T$ on $w s$ and $\mc{V}_\rho = \bigcup_{\rho' \in \mc{R}(\rho)} \mc{V}_{\rho'}$.}

			The valuation for $\sigma$ in $T$ may be written as	$\mc{V}_\sigma = \mc{V}_\rho \circ \alpha$, and $\mc{R}(\sigma)$ may be written as
			\begin{equation*}
				\mc{R}(\sigma) = \bigcup_{\beta \in \decomp{}(\alpha)} \bigcup_{\rho' \in \mc{R}(\rho)} \set{\rho' \xrightarrow[\beta]{s} \tup{r, Z} : \tup{\tup{q_n, Y_n}, s, \alpha, \tup{r, Z}} \in \Delta'}.
			\end{equation*}
			As a result, we have that 
			\begin{equation*}
				\bigcup_{\sigma' \in \mc{R}(\sigma)} \mc{V}_{\sigma'} = \bigcup_{\beta \in \decomp{}(\alpha)} \bigcup_{\rho' \in \mc{R}(\rho)} \mc{V}_{\rho'} \circ \beta.
			\end{equation*}
			Applying the inductive hypothesis and \autoref{lemma:decomp} to this equation, we obtain the desired equivalence:
			\begin{align*}
				\bigcup_{\sigma' \in \mc{R}(\sigma)} \mc{V}_{\sigma'} &= \bigcup_{\beta \in \decomp{}(\alpha)} \bigcup_{\rho' \in \mc{R}(\rho)} \mc{V}_{\rho'} \circ \beta \\
				&=\bigcup_{\beta \in \decomp{}(\alpha)} \mc{V}_\rho \circ \beta \\
				&= \mc{V}_\rho \circ \bigcup_{\beta \in \decomp{}(\alpha)} \beta \quad=\quad \mc{V}_\rho \circ \alpha \quad=\quad \mc{V}_\sigma. \qedhere
			\end{align*}
	\end{description}
\end{proof}

\begin{lem}
	\label{lemma:helper1}
	If $\tup{w, w'} \in \sem{T}$, then $\tup{w, w'} \in \sem{T'}$.
\end{lem}
\begin{proof}
	Assuming $\tup{w, w'} \in \sem{T}$, then there must be a run $\rho = q_0 \xrightarrow[\alpha_1]{w_1} \dots \xrightarrow[\alpha_n]{w_n} q_n$ in $T$ on $w$ such that $\ms{Er}_X \circ \mc{V}_\rho \circ F(q_n) = w'$.
	By \autoref{lemma:big_decomp}, it holds that
	\begin{equation*}
		w' = \ms{Er}_X \circ \bigcup_{\rho' \in \mc{R}(\rho)} \mc{V}_{\rho'} \circ F(q_n),
	\end{equation*}
	and, by \autoref{def:diamond-free_to_copyless}, it holds, for any $Y_n \subseteq X$, that
	\begin{equation*}
		w' = \ms{Er}_X \circ \bigcup_{\rho' \in \mc{R}(\rho)} \mc{V}_{\rho'} \circ F'(\tup{q_n, Y_n}),
	\end{equation*}
	such that there is no variable $x$ occurring both in $Y_n$ and in $F(q_n)$.
	Each run in $\mc{R}(\rho)$ corresponds to a nondeterministic choice resolving which copies induced by assignments in $\rho$ will be used later in the computation.
	Since each assignment in $\rho$ is diamond-free, no two copies can be combined in a later assignment, and thus there at least one run in $\mc{R}(\rho)$ corresponding to the appropriate choice of copies for simulating $\rho$.
	Therefore, $\tup{w, w'} \in \sem{T}$ implies $\tup{w, w'} \in \sem{T'}$.
\end{proof}

\begin{lem}
	\label{lemma:helper2}
	If $\tup{w, w'} \in \sem{T'}$, then $\tup{w, w'} \in \sem{T}$.
\end{lem}
\begin{proof}
	Assuming $\tup{w, w'} \in \sem{T'}$, there exists a run $\rho' = \tup{q_0, \emptyset} \xrightarrow[\beta_1]{w_1} \dots \xrightarrow[\beta_n]{w_n} \tup{q_n, Y_n}$ in $T'$ on $w$ such that 
	\begin{equation*}
		w' = \ms{Er}_X \circ \mc{V}_{\rho'} \circ F'(\tup{q_n, Y_n}).
	\end{equation*}
	By \autoref{def:diamond-free_to_copyless}, there exists a transition $\tup{q_{k-1}, w_k, \alpha_k, q_k} \in \Delta$, for each transition $\tup{\tup{q_{k-1}, Y_{k-1}}, w_k, \beta_k, \tup{q_k, Y_k}} \in \Delta'$ in $\rho'$, such that the memberships $\beta_k \in \decomp{}(\alpha_k)$ and $y \in Y_k$ hold iff the memberships $x \in \alpha_k(y)$ and $x \notin \beta_k(y)$ hold for some $x \in X$.
	Therefore, we may construct the run $\rho = q_0 \xrightarrow[\alpha_1]{w_1} \dots \xrightarrow[\alpha_n]{w_n} q_n$ for which $\rho' \in \mc{R}(\rho)$ by stringing together these transitions from $\Delta$.
	This shows that every accepting run $\rho'$ in $T'$ on $w$ has a corresponding accepting run $\rho$ in $T$ on $w$ such that 
	\begin{equation*}
		\mc{V}_\rho \circ F(q_n) = \mc{V}_{\rho'} \circ F'(\tup{q_n, Y_n}).
	\end{equation*}
	Hence, $\tup{w, w'} \in \sem{T'}$ implies $\tup{w, w'} \in \sem{T}$.
\end{proof}

\begin{thm}
	\label{theorem:diamond-free_to_copyless}
	Every diamond-free NSST with $n$ states and $m$ variables is equivalent to a copyless NSST with $n2^m$ states and $m$ variables.
\end{thm}
\begin{proof}
	Equivalence of $\sem{T}$ and $\sem{T'}$ follows immediately from the conjunction of \autoref{lemma:helper1} and \autoref{lemma:helper2}.
	The number of states and variables in $T'$ is clear from \autoref{def:diamond-free_to_copyless}.
\end{proof}

\section{Conclusion}
This paper revisits the constructions for sequential composition of copyless SSTs developed by Alur, et al. \cite{AlurCerny2010,AlurDeshmukh2011}.
While the constructions produce copyful transducers in general, we showed that the composite SSTs exhibit only the tamest of copyful behavior.
As a characterization, we defined diamond-free SSTs and proved that the composite transducer is always diamond-free.
Moreover, we formulated a method for transforming any diamond-free SST into an equivalent copyless SST.
Applying the composition construction followed by the conversion from diamond-free to copyless comprises a complete and direct procedure for composing copyless SSTs.

\bibliographystyle{alphaurl}
\bibliography{references}

\end{document}